\documentclass[11pt]{article}
\pdfoutput=1
\usepackage{jcappub,natbib}
\input{colordvi.tex}

\def\ga{\mathrel{\raise.3ex\hbox{$>$\kern-.75em\lower1ex\hbox{$\sim$}}}}
\def\la{\mathrel{\raise.3ex\hbox{$<$\kern-.75em\lower1ex\hbox{$\sim$}}}}

\title{Nonlinear perturbations from the coupling of the inflaton 
to  a non-Abelian gauge field, with a focus on 
Chromo-Natural Inflation}

\author[a]{Alexandros Papageorgiou,}
\author[a,b]{Marco Peloso,}
\author[a]{Caner  Unal}

\affiliation[a]{School of Physics and Astronomy, and Minnesota Institute for Astrophysics, University of Minnesota, Minneapolis, 55455 (USA)}

\affiliation[b]{Dipartimento di Fisica e Astronomia G. Galilei, Universit`a degli Studi di Padova,  I-35131, Padova (Italy)}

\abstract{
Several models of inflation employing a triplet of SU(2) vectors with spatially orthogonal vacuum expectation values (VEVs) have been recently proposed. One (tensor) combination $t$ of the vector modes is amplified in some momentum range during inflation. Due to the vector VEVs, this combination mixes with gravitational waves (GW) at the linear level, resulting in a GW amplification that has been well studied in the literature. Scalar perturbations in this class of models have been so far studied only at the linear level.  We perform a first step toward the nonlinear computation using as an example the original model of Chromo-Natural Inflation. We compute the contribution to the scalar power spectrum arising from the  coupling of the combination $t$ to the inflaton. This contribution is mostly controlled by a single parameter of the model (namely, the ratio between the mass of the fluctuations of the vector field and the Hubble rate), and, for a wide range of this parameter, it can significantly affect the phenomenology obtained from the linear theory. This nonlinear contribution is significantly blue, improving the comparison between the two-point function and the  Cosmic Microwave Background (CMB) data. This growth can be also relevant for smaller scale phenomenology, such as large scale structure, CMB distortions, and primordial black holes. 
}

\begin{document}

\begin{flushright}   UMN-TH 3723/18  \end{flushright}

\maketitle
\flushbottom

%%%%%%%%%%%%%%%%%%%%%%%%%%%%%%%%%%%%%
\section{Introduction}
\label{sec:intro} 
%%%%%%%%%%%%%%%%%%%%%%%%%%%%%%%%%%%%%

Axion, or natural, inflation is a class of models in which the flatness of the inflation potential is protected by an approximate shift symmetry \cite{Freese:1990rb} (see \cite{Pajer:2013fsa} for a review). In the simplest realizations, the 
inflaton has a trans-Planckian axion scale, which appears hard to reconcile with quantum gravity and string theory 
\cite{Banks:2003sx}. Among several solutions proposed for this problem \cite{ArkaniHamed:2003wu,Kim:2004rp,Dimopoulos:2005ac,Silverstein:2008sg,McAllister:2008hb,Kaloper:2008fb,Marchesano:2014mla,Bachlechner:2014gfa,Kappl:2015esy,Choi:2015aem,Parameswaran:2016qqq}, the works \cite{Anber:2009ua,Adshead:2012kp} 
considered the possibility that a sub-Planckian inflaton range can be due to the interactions of the inflaton with a gauge field. 

Ref. \cite{Anber:2009ua} studied the case of a U(1) gauge field with vanishing vacuum expectation value (vev). It was then realized that the $\chi F {\tilde F}$ interaction (where $\chi$ is the axion inflaton, $F$ the gauge field strength, and ${\tilde F}$ its dual), can lead to a very interesting phenomenology, even if it is weak enough not to significantly affect the background inflaton evolution. The motion of the inflaton amplifies one polarization of the gauge field. These modes in turn, before being diluted away by the expansion of the universe, source distinctive scalar and tensor primordial perturbations, through their nonlinear interactions  $\delta A + \delta A \rightarrow \delta \chi$ with the inflaton field and $\delta A + \delta A \rightarrow \delta g$ with the metric \cite{Barnaby:2010vf}. The phenomenological consequences of these nonlinear interactions have been well studied in the literature, including  CMB non-Gaussianity  \cite{Barnaby:2010vf,Barnaby:2011vw}, growth of the scalar power spectrum at CMB scales   \cite{Meerburg:2012id}, gravitational waves  that might be detectable by gravitational interferometers~\cite{Cook:2011hg,Barnaby:2011qe,Domcke:2016bkh,Garcia-Bellido:2016dkw,Bartolo:2016ami}, parity violation in the CMB~\cite{Sorbo:2011rz} and in interferometers~\cite{Crowder:2012ik}, primordial black holes~\cite{Linde:2012bt,Bugaev:2013fya,Erfani:2015rqv,Cheng:2015oqa,McDonough:2016xvu,Domcke:2017fix,Garcia-Bellido:2017aan}, and large and parity violating tensor bispectra~\cite{odd-TTT}.  Perturbativity limits on this coupling were studied in \cite{Ferreira:2015omg,Peloso:2016gqs}. These studies are limited to the regime of negligible backreaction of the gauge field on the inflaton dynamics. Ref.  \cite{Peloso:2016gqs} showed that perturbativity is respected in this regime. 

Ref. \cite{Adshead:2012kp} studied instead the case in which the inflaton field interacts with a SU(2) triplet of  vector fields having  nonvanishing spatial vevs. The vevs are arranged to be orthogonal to each other and of equal magnitude, so to lead to isotropic expansion. This model, dubbed ``Chromo-Natural Inflation'' shares several analogies with the model of ``Gauge-Flation''  \cite{Maleknejad:2011jw}, where a pseudo-scalar inflaton is absent and inflation is due to a $\left( F {\tilde F} \right)^2$ operator. (In fact, Gauge-Flation can be viewed as a specific limit of Chromo-Natural Inflation, in which the axion inflaton can be integrated out \cite{Adshead:2012qe,SheikhJabbari:2012qf}.) The linear theory of cosmological perturbations in Chromo-Natural Inflation was first studied in \cite{Dimastrogiovanni:2012st} in a low-energy effective description of the model, and then in \cite{Dimastrogiovanni:2012ew,Adshead:2013qp,Adshead:2013nka} in the full model. The main features emerged from these linarized studies is that the model  is unstable in a specific  regime of parameters ($m_Q < \sqrt{2}$, where $m_Q$ is introduced in eq. (\ref{mqdef})), while it is outside the allowed $n_s - r$ region in the complementary regime (where $n_s$ is the spectral tilt, and $r$ the tensor-to-scalar ratio).~\footnote{As a consequence, one should expect that also Gauge-Flation is incompatible with data, as confirmed by the analysis of \cite{Namba:2013kia}.}  

Several works modified the original model of \cite{Adshead:2012kp} so to be compatible with data, including the presence of a second axion inflaton \cite{Obata:2014loa} or a dilaton  \cite{Obata:2016tmo}, a different inflation potential \cite{Obata:2016xcr,Caldwell:2017chz,DallAgata:2018ybl}, realizations in which the axion field is not the inflaton \cite{Dimastrogiovanni:2016fuu,McDonough:2018xzh}, and a spontaneous breaking of the SU(2) symmetry \cite{Adshead:2016omu}. 

The existing phenomenological studies of these models are based on linearized perturbation theory, with the exception of 
\cite{Agrawal:2017awz,Agrawal:2018mrg,Agrawal:2018gzp} that studied the nonlinear interactions in the tensor sector, and the resulting GW bispectrum. Based on the results of the U(1) models, one could expect that nonlinearities can be of relevance also in the scalar sector. The computation in the non-Abelian context is however significantly more involved than in its U(1) counterpart: even disregarding scalar metric perturbations (which is shown to be a justified assumption \cite{Dimastrogiovanni:2012ew,Adshead:2013nka} - see Section \ref{subsec:scalar}), Chromo-Natural Inflation has three scalar perturbations coupled to each other at the linearized level;  this set comprises of the inflation perturbation plus two linear combinations of perturbations of the gauge fields.  For this reason already the linearized computation is significantly more involved in the non-Abelian vs. the Abelian case, and we expect this to be true also at the nonlinear level.

\begin{figure}[tbp]
\centering 
\includegraphics[width=0.6\textwidth,angle=0]{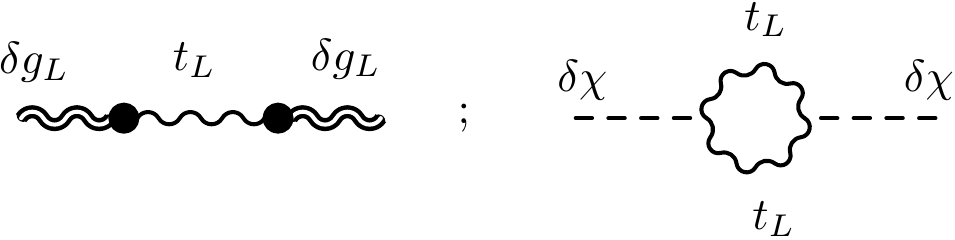}
\caption{Diagramatic representation of the GW (left diagram) and scalar (right diagram) power spectra sourced by the enhanced tensor mode $t_L$. While in the non-Abelian models the GW are sourced already at the linearized level, the scalar modes are sourced at the nonlinear level. The right diagram is the interaction studied in this work. 
}
\label{fig:diagrams}
\end{figure}

With this in mind, in the present work we only perform a first step toward the full nonlinear computation. Specifically, we consider one nonlinear interaction of the inflaton perturbation; this mode is the dominant scalar perturbation in the super-horizon regime, where it coincides (up to negligible corrections) with the adiabatic mode  $\zeta \simeq - \frac{H}{\dot{\chi}} \, \delta \chi$. In the U(1) case, the motion of the inflaton significantly amplifies one gauge field polarization at horizon crossing. 
In the present context, the background dynamics amplifies one polarization of one (tensor) combination $t$. For definiteness, let us assume that the amplified polarization is the left-handed one.~\footnote{Changing the sign of the $\frac{\lambda}{f} \chi F {\tilde F}$ term, or of $\dot{\chi}$, switches the role of the two polarizations.} The mode $t_L$ plays an analogous role to the mode $\delta A_L$ in the Abelian case. In the U(1) case, the amplification is exponentially sensitive to the parameter $\xi \equiv \frac{\lambda \dot{\chi}}{2 f H}$ (where $H$ is the Hubble rate), while in the present context the amplification is exponentially sensitive to the parameter $m_Q$ defined in eq. (\ref{mqdef}).~\footnote{We note from eq. (\ref{slowroll}) that $m_Q$ coincides with $\xi$ in the large $m_Q$ regime.}  In the present context, $t_L$ is coupled to the GW at the linearized level, sourcing the GW mode $h_L$ through the left digram in Figure \ref{fig:diagrams}.~\footnote{Refs. \cite{Caldwell:2016sut,Caldwell:2017sto} studied in a more general context the mixing, and the resulting oscillations, between GW and gauge modes in a stationary gauge field background.}  The generation of this chiral GW background is probably the most distinctive phenomenological feature obtained so far for this class of models  \cite{Adshead:2013qp,Maleknejad:2016qjz,Agrawal:2017awz,Caldwell:2017chz,Thorne:2017jft,Agrawal:2018mrg}. This linear coupling is absent in the U(1) case, where a chiral GW background is produced by the $\delta A_L + \delta A_L \rightarrow \delta g_L$ process \cite{Barnaby:2010vf,Sorbo:2011rz}. Both in the Abelian, and non-Abelian case, the amplification of $t_L$ affects the scalar modes only at the nonlinear level. In this work we compute the right diagram of figure \ref{fig:diagrams}, which is the direct counterpart of the $\delta A_L + \delta A_L \rightarrow \delta \chi$ process present in the U(1) case. Even if it is only one of the diagrams contributing to the nonlinear scalar power spectrum, the fact that this interaction is much stronger than the gravitational ones (see Section~\ref{sec:nonlinear}), that $t_L$ is the enhanced mode, and that $\delta \chi $ is mapped into the observed scalar perturbation induces us to argue that  its value can be representative of the complete one.
 The results obtained in this work show that this contribution can be sizable for a significant range of parameters, and it can significantly affect the phenomenology of these models, due to its strong scale dependence. This serves as a motivation for a more complete analysis, in which all the scalar perturbations are included. 

The plan of this paper is the following. In Section \ref{sec:cn} we review the model of Chromo-Natural Inflation, and provide a novel analytic approximation for the background evolution. In Section \ref{sec:linear} we summarize the results obtained from the linearized study of the perturbations. In Section \ref{sec:nonlinear} we present the computation of the second diagram in Figure \ref{fig:diagrams}, and discuss how this can affect the phenomenology of the model.  In Section \ref{sec:conclusions} we present our conclusions, with prospects for future work. The paper is concluded by three Appendices. In Appendix \ref{app:A} we provide the solutions of the scalar perturbations in the UV regime, as necessary to provide their initial conditions; in Appendix \ref{app:B} we present details of the diagrammatic computation; in Appendix \ref{sec:appC} we perform a semi-analytical study of the result, to understand its scaling with the parameter $m_Q$.

%%%%%%%%%%%%%%%%%%%%%%%%%%%%%%%%%%%%%
\section{Chromo-Natural Inflation} 
\label{sec:cn} 
%%%%%%%%%%%%%%%%%%%%%%%%%%%%%%%%%%%%%

In this section, we give a brief summary of the original model of Chromo-Natural Inflation \cite{Adshead:2012kp}.  It is a model in which the slow evolution of the inflaton is due to its interactions with an SU(2) vector field having a nonvanishing spatial vacuum expectation value (vev), 
\begin{equation}
\left\langle A^a_0 \left( t \right) \right\rangle =0, \quad \left\langle A^a_i \left( t \right) \right\rangle = \delta^a_i \, a \left( t \right) \,  Q \left( t \right)  \;. 
\end{equation}
In this expression, $a = \left\{ 1 ,\, 2 ,\, 3 \right\}$ is the SU(2) index, while $a \left( t \right)$ is the scale factor (we choose the line element as $ds^2 = - d t^2 + a^2 \left( t \right) d \vec{x}^2 = a^2 \left( \tau \right) \left[ - d \tau^2 + d \vec{x}^2 \right]$). The indices $0$ and $i= \left\{ 1 ,\, 2 ,\, 3 \right\}$ are time and space indices, respectively. The vev is chosen so to be compatible with a homogeneous and isotropic expansion. 
The vector vev is parametrized as $a \left( t \right) Q \left( t \right)$ since, as we will see, $Q$ is slowly evolving during inflation. 

The lagrangian of the model is
\begin{equation}
S= \int d^4x \sqrt{-g} \left[ \frac{M_p^2}{2}R \, -\frac{1}{4}F^a_{\mu \nu}F^{a , \mu \nu} -\frac{1}{2}\left(\partial \chi \right)^2 -V(\chi)  - \frac{\lambda}{8 \sqrt{-g} \, f} \chi \, \epsilon^{\mu \nu \alpha \beta} F^a_{\mu \nu}F^{a}_{\alpha \beta}  \right] \;, 
\end{equation}
where $\epsilon^{\mu \nu \alpha \beta}$ is totally anti-symmetric, and normalized to $\epsilon^{0123}=1$, $\chi$ is the pseudo-scalar inflaton, and F is the field strength of the SU(2) field, 
$F^a_{\mu \nu}=\partial_\mu \, A^a_\nu - \partial_\nu \, A^a_\mu  + g \, \epsilon^{abc} A^b_\mu \, A^c_\nu \,$. 

Assuming that the inflaton has a homogeneous and time dependent vev, one obtains the background equations 
\begin{eqnarray}
&& 3H^2 M_p^2  = \frac{1}{2}\dot \chi^2 + V(\chi) + \frac{3}{2} \left[  \left( \dot Q + HQ \right)^2+ g^2 Q^4  \right] \;, \nonumber\\
&& \ddot \chi + 3 H \dot \chi + V'(\chi)  +  \frac{3 \,\lambda \, g}{f} \,Q^2  \left( \dot Q  + H Q  \right) =0 \;, \nonumber\\
&& \ddot Q + 3 H \dot Q + (\dot H +2 H^2) Q + g\, Q^2 \left( 2 g Q - \frac{\lambda \dot \chi}{f} \right) = 0 \;, 
\end{eqnarray}
where dot denotes a time derivative, and $H$ is the Hubble rate. The slow roll solution (neglecting $\ddot \chi, \ddot Q$ and $\dot H$) with  strong ``magnetic drag force''  ($3 f^2 \, H^2 \ll g^2 \lambda^2 \, Q^4$ and $\lambda^2 \, Q^2 \gg 2 f^2$) gives 
\begin{eqnarray}
\dot Q \simeq -H Q -\frac{f \, V'(\chi)}{3 \,g \, \lambda \, Q^2} \;\;,\;\; 
\dot \chi  \simeq  \frac{f \, H}{g\, \lambda \, Q^2} \left( \frac{2 g^2 Q^3}{H}  - H Q - \frac{f \, V'(\chi)} {g \lambda Q^2} \right) \;. 
\label{qchisreq}
\end{eqnarray}

One can verify \cite{Adshead:2012kp,Dimastrogiovanni:2012ew} that the left hand side can be neglected in the first of these relations, so that 
\begin{equation}
Q \simeq \left(  - \frac{f \, V'(\chi)}{3 g \lambda H} \right)^{1/3}  \;\; \Rightarrow \;\; 
\dot \chi \simeq \, 2 \, \frac{f}{\lambda} \, H \, \left( \frac{g \, Q}{H} + \frac{H}{g \, Q} \right) =  2 \, \frac{f}{\lambda} \, H \left( m_Q + \frac{1}{m_Q} \right) \;, 
\label{slowroll}
\end{equation}
where in the last expression we have introduced the dimensionless parameter 
\begin{equation}
m_Q \equiv \frac{g \, Q}{H} \;\;. 
\label{mqdef}
\end{equation}

It is useful to obtain a relation between the value of $\chi$ and the number of e-folds in this model. To do so, we follow   \cite{Adshead:2012kp} in choosing the simplest potential for the axion inflaton, 
\begin{equation}
V(\chi)= \mu^4 \left[ 1+ \cos \, {\tilde x} \right] \;\;\;,\;\;\; {\tilde x} \equiv \frac{\chi}{f}  \;, 
\label{potential} 
\end{equation}
and assume that the inflaton rolls in the region between the maximum at $x=0$ and the minimum $x = \pi$. We  integrate the relation $dN  =H \, dt= \frac{H}{\dot \chi} \, d\chi $ to obtain 
\begin{equation}
N  =  \frac{\lambda \sqrt{y}}{2} \, \int_{{\tilde x}_N}^{{\tilde x}_{\rm end}} \, d{\tilde x} \, \frac{F \left( {\tilde x} \right)}{1 + y \, F^2 \left( {\tilde x} \right)} \;\;,\;\; 
F \left( {\tilde x} \right) \equiv \frac{\left( 1 + \cos {\tilde x} \right)^{2/3}}{\sin^{1/3} \, {\tilde x}}  \;, 
\label{N-chiN-F}
\end{equation}
where we have introduced the combination $y \equiv \left( \frac{\lambda \tilde \mu^4} {3 g^2}  \right)^{2/3}$, with 
${\tilde \mu } \equiv \frac{\mu}{M_p}$. The quantity ${\tilde x}_{\rm end}$ is the (rescaled) value of the inflaton at the end of inflation. With very good approximation, inflation ends at the minimum of the potential, so we set ${\tilde x}_{\rm end} = \pi$. The quantity ${\tilde x}_N$ is the (rescaled) value of the inflaton at $N$ e-folds before the end of inflation. 

The relations presented so far were derived in  \cite{Adshead:2012kp} (see also \cite{Dimastrogiovanni:2012ew}) where the reader is referred to for more details. Here we point out that the relation (\ref{N-chiN-F}) admits a simple approximate solution in the regime that is phenomenologically relevant. Specifically, we can linearize the function $F$ next to the minimum of the potential, $F \simeq \frac{\pi-{\tilde x}}{2^{2/3}}$. This relation is exact for ${\tilde x} = \pi$, and it approximates the exact one with $1\%$ accuracy at ${\tilde x} = \frac{\pi}{2}$. Using this approximate expression, the relation (\ref{N-chiN-F}) gives 
\begin{equation}
N \simeq \frac{\lambda}{2^{4/3} \sqrt{y}} \, \ln \left[ 1 + \frac{y \left( \pi - {\tilde x}_N \right)^2}{2^{4/3}} \right] \simeq \frac{\lambda \, \sqrt{y} \left( \pi - {\tilde x}_N \right)^2}{2^{8/3}} \;\; \Rightarrow \;\; {\tilde x}_N \simeq \pi - \frac{2^{4/3} \, \sqrt{N}}{y^{1/4} \, \sqrt{\lambda}}  \;, 
\label{xN}
\end{equation} 
where we have further assumed that the argument in the square bracket is close to $1$. The requirement that this is true, and that ${\tilde x}_N$ is greater than $\frac{\pi}{2}$ (so that the linearized approximation for $F$ is accurate), translate, respectively, into the bounds $N \ll \frac{0.4 \, \lambda}{\sqrt{y}}$ and $N < 0.4 \sqrt{y} \, \lambda $. These bounds are compatible with $N=60$ e-folds of inflation, provided that $\lambda$ is sufficiently large (it is known that $\lambda \gg 1$ is required in this model to have a sufficiently long inflation \cite{Adshead:2012kp}). 

Inserting the expression of the potential in the slow roll relation (\ref{slowroll}) for $Q$, and in $H \simeq \frac{\sqrt{V}}{\sqrt{3} M_p}$, we obtain 
\begin{equation}
m_Q \left( N \right) \simeq \frac{1}{\sqrt{y} \, F \left( {\tilde x}_N \right)} \simeq \frac{\sqrt{\lambda}}{2^{4/3} \, y^{1/4}} \, \frac{1}{\sqrt{N}} 
\simeq 0.098 \left( \frac{\lambda \, g}{{\tilde \mu}^2} \right)^{1/3} \, \sqrt{\frac{60}{N}} \;. 
\label{mQ-N}
\end{equation} 
As we shall see, the quantity $m_Q$ is the key combination that controls the perturbations in this model. This simple analytical result will allow us to gain a better understanding of how our results scale with the parameters in the model. Moreover, as it is standard in inflation, perturbations probe the background evolution at horizon crossing; therefore knowing how $m_Q$ evolves with $N$ will allow us to infer the scale dependence of the perturbations of the model.

\begin{figure}[tbp]
\centering 
\includegraphics[width=0.6\textwidth,angle=0]{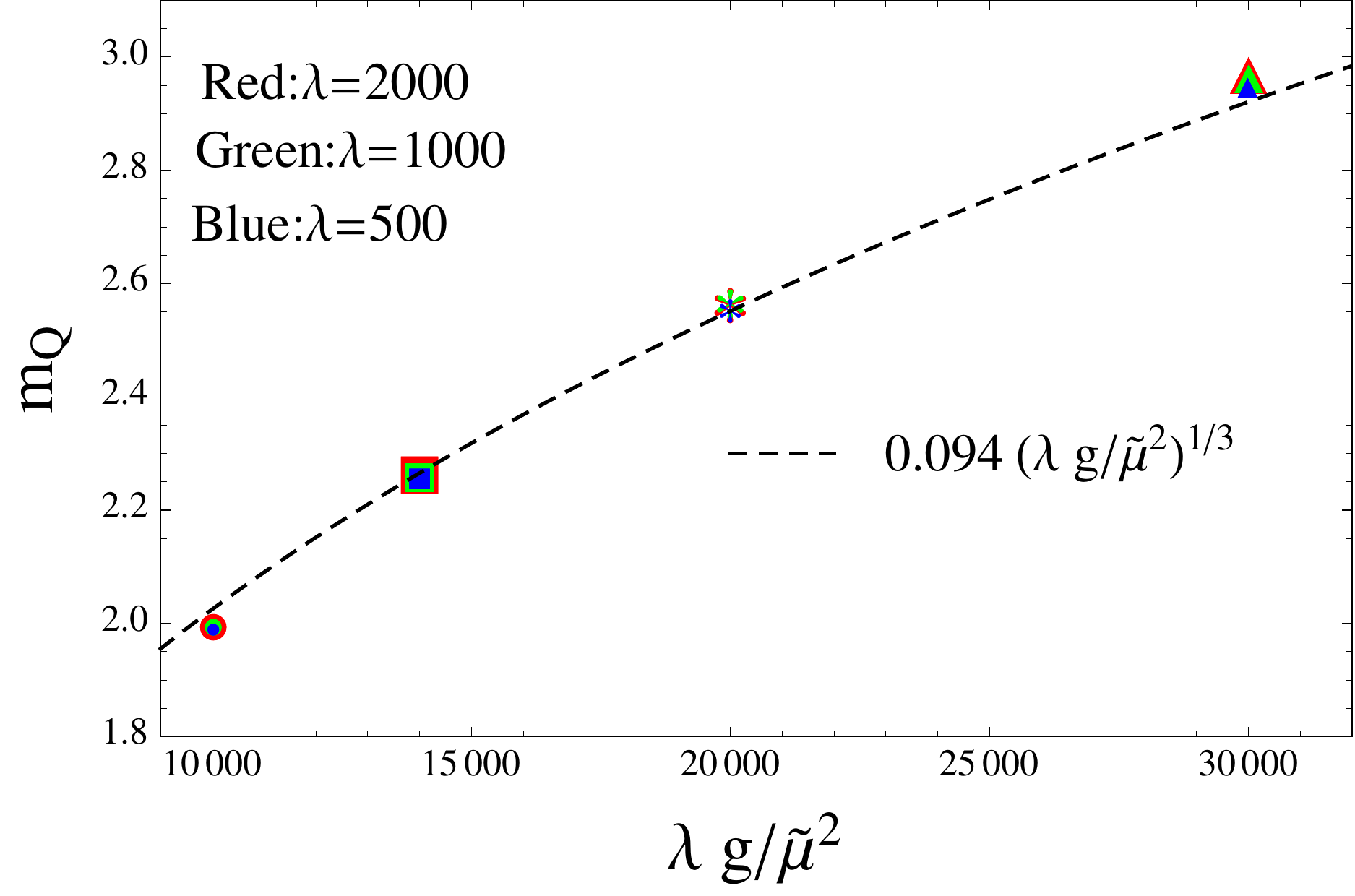}
\caption{Value of $m_Q$ at $N=60$ e-folds of inflation, for different choices of parameters described in the main text.  Points with different $\lambda$, but equal $\lambda g / {\tilde \mu}^2$ give nearly overlapping results in the plot. Also shown is the analytical result (\ref{mQ-N}), with a $\sim 4\%$ change of the numerical coefficient to better fit the numerical results. } 
\label{fig:mQ}
\end{figure}

With the choice (\ref{potential}) of the potential, the model is characterized by the $4$ parameters, $\lambda ,\,  {\tilde f} \equiv \frac{f}{M_p} ,\, g$, and ${\tilde \mu} \equiv \frac{\mu}{M_p}$. A direct inspection of the equations of motion shows that the last two parameters affect the  background evolution (and so $m_Q$) only through the ratio $\frac{g}{{\tilde \mu}^2}$. This degeneracy is broken only when we fix the scale of the potential ${\mu}$ through the amplitude of the scalar perturbations. To verify the accuracy of (\ref{mQ-N}), we performed several 
background evolutions of the model, characterized by $\lambda = 500$ and  $\frac{g}{{\tilde \mu}^2} = \left\{ 20 ,\, 28 ,\, 40 ,\, 60 \right\}$; then by $\lambda = 1,000$ and  $\frac{g}{{\tilde \mu}^2} = \left\{ 10 ,\, 14 ,\, 20 ,\, 30 \right\}$; then by $\lambda = 2,000$ and  $\frac{g}{{\tilde \mu}^2} = \left\{ 5 ,\, 7 ,\, 10 ,\, 15 \right\}$. In all cases we fixed ${\tilde f} = 0.1$. The numerical value obtained by $m_Q$ at $N= 60$ e-folds was then compared with the final expression in eq. (\ref{mQ-N}), with a slight change of the numerical coefficient to better fit the data. We see from Figure \ref{fig:mQ} that the relation (\ref{mQ-N}) is indeed extremely accurate in all these cases. We have verified that the same holds true for ${\tilde f} = 0.01$ (not shown here).

%%%%%%%%%%%%%%%%%%%%%%%%%%%%%%%%%%%%%
\section{Linear theory of the perturbations} 
\label{sec:linear} 
%%%%%%%%%%%%%%%%%%%%%%%%%%%%%%%%%%%%%

In this section we describe the linear perturbations in the model of Chromo-Natural Inflation. Due to the presence of vector vevs, some perturbations of the gauge fields mix with tensor perturbations of the metric at the linearized level. Following the convention used in the literature, we collectively denote these modes as ``tensor perturbations''.  These perturbations further split in two subsets, that are characterized by opposite helicity, and that are decoupled at the linearized level. The two subsets give different results, due to the parity-violating nature of the model. Most interestingly, the tensor modes in the gauge field of one given helicity manifest a tachyonic growth 
close to horizon crossing. This mode sources tensor metric perturbations of that helicity. The production of an enhanced chiral GW background is one of the most interesting phenomenological outcome of this class of models. 

Also due to the vector vevs, some other perturbations of the gauge fields are coupled at the linearized level with the perturbation of the inflaton and with the scalar perturbations of the metric. These modes are collectively denoted as ``scalar perturbations''. At the linearized level, the scalar sector does not present an enhancement analogous to that of the tensor perturbations, provided that the parameter $m_Q$ defined in eq. (\ref{mqdef}) satisfies $m_Q > \sqrt{2}$ (in the opposite regime, the scalar perturbations make the background solution considered in the previous section unstable). In the next section we argue that this conclusion does not necessarily hold when nonlinear interactions between the enhanced tensor mode and the inflaton perturbation are taken into account. 

This section is divided in three parts. In the first two parts we review, respectively, the results for the tensor and scalar modes obtained in the literature. We then give a brief summary of the phenomenology resulting by this linearized computations. 

We use the convention 
\begin{equation}
\delta \left( t ,\, \vec{x} \right) = \int \frac{d^3 k}{\left( 2 \pi \right)^{3/2}} \, {\rm e}^{i \vec{k} \cdot \vec{x}} \, \delta \left( t ,\, \vec{k} \right) \;, 
\label{FT} 
\end{equation} 
for the Fourier transform of any perturbation $\delta$.

%%%%%%%%%%%%%%%%%%%%%%%%%%%%%%%%%%%%%
\subsection{Tensor sector} 
\label{subsec:tensor} 
%%%%%%%%%%%%%%%%%%%%%%%%%%%%%%%%%%%%%

Tensor perturbations in this model were studied first in \cite{Dimastrogiovanni:2012ew} and then, with a better accuracy, in \cite{Adshead:2013qp}. In the linearized computation described here, different perturbations are not coupled to each other. Therefore, without loss of generality, we can orient the momentum of the modes along the $z-$axis. Doing so, the tensor perturbations of the model can be written as 
\begin{eqnarray} 
&& \delta A_\mu^1 = a \left( \tau \right) \left( 0 ,\, t_+ \left( \tau ,\, z \right) ,\, t_\times \left( \tau ,\, z \right) ,\, 0 \right) \;\;,\;\; 
\delta A_\mu^2 = a \left( \tau \right) \left( 0 ,\, t_\times \left( \tau ,\, z \right) ,\, - t_+ \left( \tau ,\, z \right) ,\, 0 \right) \;\;, \nonumber\\ 
&& \delta g_{11} = - \delta g_{22} = \frac{a^2 \left( \tau \right)}{\sqrt{2}} h_+ \left( \tau ,\, z \right) \;\;,\;\; 
\delta g_{12} = \delta g_{21} = \frac{a^2 \left( \tau \right)}{\sqrt{2}} h_\times \left( \tau ,\, z \right) \;, 
\label{tensor-modes}
\end{eqnarray} 
where $\tau$ is conformal time. Starting from these modes, we define the left handed (+) and right handed (-) helicity variables 
\begin{equation}
{\hat h}^{\pm} \equiv \frac{a M_p}{2} \frac{ h_+  \mp i h_\times }{\sqrt{2}} \equiv \frac{a M_p}{2} h_{L,R}  \;\;\;\;\;\;,\;\;\;\;\;\; 
{\hat t}^{\pm} \equiv \sqrt{2} a \, \frac{ t_+  \mp i t_\times }{\sqrt{2}} \equiv \sqrt{2} a \, t_{L,R}  \;. 
\label{tensor-canonical}
\end{equation} 
These modes are canonically normalized (namely, their kinetic action in momentum space is $S_{\rm kin} = \frac{1}{2} \int d\tau d^3 k \left[ 
\vert {\hat t}^{+'} \vert^2 + \vert {\hat h}^{+'} \vert^2 + \vert {\hat t}^{-'} \vert^2 + \vert {\hat h}^{-'} \vert^2 \right]$). Moreover, the two subset of modes $\left\{ {\hat t}^+ ,\, {\hat h}^+ \right\}$ and $\left\{ {\hat t}^- ,\, {\hat h}^- \right\}$ are decoupled from each other at the linearized level. 

Disregarding their coupling with the metric perturbations (which can be verified to be a good approximation a posteriori in the left handed helicity sector, which is the one of interest for the present discussion) the two tensor modes of the gauge field satisfy \cite{Adshead:2013qp}
\begin{equation} 
\frac{d^2}{d x^2} \, {\hat t}^{\pm} + \left( 1 + \frac{m}{x^2} \mp \frac{m_t}{x} \right) t^{\pm} = 0 \;\;,\;\; x \equiv - k \tau \;, 
\label{eq-t}
\end{equation} 
where we have defined the two positive quantities 
\begin{equation} 
m \equiv 2 \left( 1 + m_Q^2 \right) \equiv \frac{1}{4} - \beta^2 \;\;,\;\; 
m_t \equiv 2 \left( 2 m_Q + \frac{1}{m_Q} \right) \equiv -2 i \alpha \;. 
\end{equation} 

As we will see later, $m_Q = {\rm O } \left( 1 \right)$ in the regime of phenomenological interest. Therefore the left handed ${\hat t}^+$ mode  has a tachyonic mass for some finite time close to horizon crossing. We therefore concentrate our discussion on this mode. The quantity $m_Q$ has a slow roll variation in this expression, see eq. (\ref{mQ-N}). As done in  \cite{Adshead:2013qp} we treat this as a constant. In studying the spectral dependence, one can evaluate $m_Q$ at the scale at which any given mode leaves the horizon (as it is customarily done in models of slow roll inflation). In this limit, one obtains the analytic solution  \cite{Adshead:2013qp} 
\begin{equation}
{\hat t}^+ = A_k \, M_{\alpha,\, \beta} \left( 2 i x \right) + B_k \, W_{\alpha,\beta} \left( 2 i x \right) \;, 
\label{exacttmode}
\end{equation} 
where $M$ and $W$ are Whittaker functions, and where the two integration constants are 
\begin{eqnarray} 
A_k = \frac{1}{\sqrt{2k}}\frac{\Gamma\left(-\alpha+\beta+\frac{1}{2}\right)}{(2i)^{-\alpha}\Gamma(2\beta+1)} \;\;,\;\; 
B_k = - \frac{1}{\sqrt{2k}} \frac{\Gamma\left(-\alpha+\beta+\frac{1}{2}\right)}{\Gamma\left(\alpha+\beta+\frac{1}{2}\right)}2^\alpha i^{\beta+1}(-i)^{\alpha-\beta}  \;. 
\end{eqnarray} 
This choice gives 

\begin{equation}
\lim_{x \to \infty}  A_k \, M_{\alpha,\beta} \left( 2 i x \right) + B_k \, W_{\alpha,\beta} \left( 2 i x \right) = \frac{{\rm e}^{i x - \alpha \, \ln x}}{\sqrt{2 k}} \;, 
\end{equation} 
which is the adiabatic vacuum solution (we note the presence of the slowly evolving factor proportional to $\alpha$, in addition to the standard adiabatic vacuum solution obtained in most models of inflation; this is due to a $\propto {\hat t} \, {\hat t}'$ term in the linearized action for the tensor modes). 

The evolution of ${\hat t}^+$ is further discussed in Appendix \ref{sec:appC}.

Next, we consider the equation for the left-handed metric perturbation in the presence of the ${\hat t}^+$ mode. To leading order in slow roll, it reads 
\begin{equation}
\frac{d^2}{d x^2} {\hat h}^+ + \left[ 1 - \frac{2}{x^2}  \right] {\hat h}^+ = 2 \frac{Q}{M_p x} {\hat t}^{+'} + 2 m_Q \left( m_Q - x \right) \frac{Q}{M_p x^2} {\hat t}^+  \;. 
\label{eq-h+}
\end{equation} 

Ref.  \cite{Adshead:2013qp} provided the approximate solution of this equation. The power in the positive helicity GW mode is given by 
\begin{equation} 
P_L = \frac{k^2}{2 \pi^2 } \, \left\vert \frac{2}{a M_p} \, {\hat h}^+ \right\vert^2 \simeq \frac{H^2}{\pi^2 M_p^2}  \left\{ 1 + 2 \, k B_k^2 \, \frac{Q^2}{M_p^2} \, \left[  \frac{\pi   \, {\cal A}   }{ \cos \left( \beta \pi \right)\left( 9 - 40 \beta^2+ 16 \beta^4 \right)}  \right]^2 \right\} \;, 
\label{PL}
\end{equation} 
where 
\begin{eqnarray}
{\cal A} &\equiv& 
\frac{1}{ \Gamma \left( \frac{1}{2} - \alpha - \beta \right) \Gamma \left( \frac{1}{2} - \alpha + \beta \right)} 
 \Bigg\{ 
 \left( i + m_Q \right) \left( 9 - 40 \beta^2 + 16 \beta^4 \right) \Gamma \left( - \alpha \right) \nonumber\\ 
&& - 8 i \left[ 2 + 16 \alpha - 8 \beta^2 + m_Q \left( - 9 i + m_Q + 8 m_Q \alpha - 4 \left( - i + m_Q \right) \beta^2 \right) \right] \Gamma \left( 1 - \alpha \right)  
 \Bigg\}  \nonumber\\ 
&& - \frac{ i }{ \Gamma \left( 1 - \alpha \right) } 
\Bigg\{  
9 - 40 \beta^2 + 8 m_Q^2 \alpha \left( - 1 + 8 \alpha + 4 \beta^2 \right) \nonumber\\ 
& &  + i m_Q \left( - 9 + 4 \beta^2 \right) \left( - 1 + 8 \alpha + 4 \beta^2 \right) 
+ 16 \left[ \alpha \left( - 1 + 8 \alpha \right) + 4 \alpha \beta^2 + \beta^4 \right] \Bigg\} \;. 
\end{eqnarray}
The two contributions originate, respectively, from the homogeneous and the particular solution of eq. (\ref{eq-h+}). The two terms are statistically uncorrelated, and therefore their powers add up without interference. For the right helicity mode the contribution from ${\hat t}^-$ can be disregarded (as this mode does not experience tachyonic growth), and we have the standard result 
\begin{equation} 
P_R = \frac{k^2}{2 \pi^2 } \, \left\vert \frac{2}{a M_p} \, {\hat h}^- \right\vert^2 \simeq \frac{H^2}{\pi^2 M_p^2} \;. 
\label{PR}
\end{equation} 
%

%%%%%%%%%%%%%%%%%%%%%%%%%%%%%%%%%%%%%
\subsection{Scalar sector } 
\label{subsec:scalar} 
%%%%%%%%%%%%%%%%%%%%%%%%%%%%%%%%%%%%%

The linear scalar perturbations in this model were first computed in \cite{Dimastrogiovanni:2012ew}, where it was shown that 
the scalar sector is stable provided that $m_Q > \sqrt{2}$.  This regime was previously studied in \cite{Dimastrogiovanni:2012st} 
in an effective single field low-energy description (based on the gelaton mechanism).  The study of \cite{Dimastrogiovanni:2012ew} is based on exact numerical computations, supplemented by analytical WKB solutions in the sub-horizon regime. A further study of the scalar perturbations, with a more extensive analysis of the phenomenology of the model, was then performed in \cite{Adshead:2013qp,Adshead:2013nka}. As proven in \cite{Dimastrogiovanni:2012ew,Adshead:2013nka}, scalar metric perturbations can be disregarded to leading order in slow roll (more accurately, we can work in spatially flat gauge, $\delta g_{ij,{\rm scalar}} = 0$. In this gauge, the scalar metric perturbations contain two non-dynamical modes. Integrating these modes out provides negligible contributions to the equations of the dynamical modes
 \cite{Dimastrogiovanni:2012ew,Adshead:2013nka}.) 

One is therefore left with the scalar perturbations in the inflaton field $\delta \chi \left( t ,\, z \right)$, and in the SU(2) multiplet, 
\begin{eqnarray}
\delta A_\mu^1 &=& \left( 0 ,\, \delta \phi \left( t ,\, z \right) - Z \left( t ,\, z \right) ,\, \chi_3 \left( t ,\, z \right) ,\, 0 \right) \;, \nonumber\\ 
\delta A_\mu^2 &=& \left( 0 ,\, - \chi_3 \left( t ,\, z \right) , \delta \phi \left( t ,\, z \right) - Z ,\, 0 \right) \;, \nonumber\\ 
\delta A_\mu^3 &=& \left( \delta A_0^3 \left( t ,\, z \right) ,\, 0 ,\, 0 ,\, \delta \phi + 2 Z \left( t ,\, z \right) \right) \;, 
\end{eqnarray} 
together with the gauge choice 
\begin{equation}
\chi_3 = - i k \, \frac{2 Z + \delta \phi}{2 g a Q} \;. 
\label{gauge-chi3}
\end{equation} 
We have adopted the scalar decomposition made in \cite{Adshead:2013nka}, with the momentum of the modes oriented along the $z-$axis (as discussed in the previous subsection, we can do so without loss of generality in a linearized computation). The mode $\delta A_0^3 $ is nondynamical, and its equation of motion is a constraint equation that we can solve to express this mode as a function of the remaining scalar perturbations. Using (\ref{gauge-chi3}), this constraint equation gives~
\begin{equation}
\delta A_0^3 = - \frac{4 g \left( a \, Q \right)' \chi_3 - g a^2 Q^2 \frac{\lambda}{f} \left( - i k \right) \delta \chi}{k^2 + 2 g^2 a^2 Q^2} \,. 
\label{constraint} 
\end{equation}

Once the two modes $\chi_3$ and $\delta A_0^3$ are eliminated through (\ref{gauge-chi3}) and (\ref{constraint}) we have a system containing the three dynamical scalar perturbations $\delta \chi ,\, \delta \phi,$ and $Z$. In terms of the canonical modes 
\begin{equation}
{\hat X} \equiv a \, \delta \chi \;\;,\;\; {\hat Z} = \sqrt{2} \left( Z - \delta \phi \right) \;\;,\;\; 
{\hat \varphi} \equiv \sqrt{2 + \frac{x^2}{m_Q^2}} \left( \frac{\delta \phi}{\sqrt{2}} + \sqrt{2} \, Z \right) \;\;, 
\label{scalar-canonical}
\end{equation} 
the equations for the dynamical scalar perturbations read 
\begin{eqnarray} 
&& {\hat X}'' + \left( 1 - \frac{2}{x^2} + \frac{V''}{H^2 x^2} + \frac{\Lambda^2 m_Q^2}{2 m_Q^2 + x^2} \right) {\hat X} 
+ \left( \frac{\chi^{' 2}}{2 M_p^2} + \frac{m_Q^2 \, Q^2}{x^2 \, M_p^2} + \frac{Q^2}{x^2 M_p^2} \right) {\hat X} 
\nonumber\\ 
&& + \frac{{\bf \Lambda} m_Q \sqrt{4+\frac{2 x^2}{m_Q^2}} \left( 4 m_Q^4 + 3 m_Q^2 x^2 + x^4 \right)}{\left( 2 m_Q^2 x + x^3 \right)^2} {\hat \varphi} - \frac{2 \sqrt{2} {\bf \Lambda} m_Q}{x^2} {\hat Z} - \frac{{\bf \Lambda} m_Q}{x \sqrt{1 + \frac{x^2}{2 m_Q^2}}} {\hat \varphi}' + 
\frac{\sqrt{2} {\bf \Lambda} m_Q}{x} {\hat Z}' = 0 \;, \nonumber\\ 
& &  {\hat \varphi}'' + \left( 1 - \frac{2}{2 m_Q^2 + x^2} + \frac{2 m_Q^2}{x^2} + \frac{6 m_Q^2}{\left( 2 m_Q^2 + x^2 \right)^2} \right) {\varphi} 
+ \frac{2 \sqrt{2 + \frac{x^2}{m_Q^2}}}{x^2} {\hat Z} + \frac{{\bf \Lambda} m_Q}{x \sqrt{1+\frac{x^2}{2 m_Q^2}}} \, {\hat X}' \nonumber\\ 
& & + \frac{{\bf \Lambda} m_Q \sqrt{4 + \frac{2 x^2}{m_Q^2}} \left( 2 m_Q^2 + m_Q^2 x^2 + x^4 \right)}{\left( 2 m_Q^2 x + x^3 \right)^2} {\hat X} = 0 \;, 
\nonumber\\ 
& & {\hat Z}'' + \left( 1 - \frac{2 -2 m_Q^2}{x^2} \right) {\hat Z} - \frac{\sqrt{2} {\bf \Lambda} m_Q}{x^2} \left( {\hat X} + x {\hat X}' \right) 
+ \frac{2 \sqrt{2 + \frac{x^2}{m_Q^2}}}{x^2} {\hat \phi} = 0 \;, 
\label{4.11-4.12-4.13-us}
\end{eqnarray} 
where ${\bf \Lambda} \equiv \frac{\lambda}{f} Q$. Using eqs. (\ref{xN}) and (\ref{mQ-N}), we can approximate 
\begin{equation}
{\bf \Lambda} \simeq \frac{5.4 \cdot 10^{-4} \, \lambda^2}{{\tilde f} m_Q^2 \left( N/60 \right)^{3/2}} \sqrt{1-\cos \left( \frac{240 m_Q \left( N/60 \right)}{\lambda}  \right)} \simeq \frac{0.09 \,\lambda}{{\tilde f} m_Q \sqrt{N/60}} \;, 
\label{Lambda-scaling}
\end{equation} 
where the last expression holds for small argument in the cosine. Prime in the expressions (\ref{4.11-4.12-4.13-us}) denotes a derivative with respect to $x \equiv - k \tau$.  The last two equations agree with equations (4.12) and (4.13) of \cite{Adshead:2013nka}. The first equation has an additional term with respect to (4.11) of \cite{Adshead:2013nka}, given by the last parenthesis in the first line. These terms are slow roll suppressed with respect to those entering in the previous parenthesis, which is probably the reason why they are not present in  \cite{Adshead:2013nka}. We ignore them in our numerical evolution. 

As we show in Appendix \ref{app:A}, these equations admit the adiabatic early time (large $x$) vacuum solution  \cite{Adshead:2013nka} 
\begin{eqnarray}
{\hat X}_{\rm in} &=&  \frac{\sqrt{1+m_Q^2}}{\sqrt{2 k}}  \,  {\rm e}^{i \left( x - x_{\rm in} \right)} \, \left( \frac{x_{\rm in}}{x} \right)^{i \sqrt{\frac{1+m_Q^2}{2}} \, {\bf \Lambda}} \;, \nonumber\\ 
{\hat \varphi}_{\rm in} &=& -  \frac{1}{\sqrt{2 k}}  \,   {\rm e}^{i \left( x - x_{\rm in} \right)} \, \left( \frac{x_{\rm in}}{x} \right)^{i \sqrt{\frac{1+m_Q^2}{2}} \, {\bf \Lambda}} \;, \nonumber\\ 
{\hat Z}_{\rm in} &=&    \frac{i  m_Q}{\sqrt{2 k}} \,   {\rm e}^{i \left( x - x_{\rm in} \right)} \, \left( \frac{x_{\rm in}}{x} \right)^{i \sqrt{\frac{1+m_Q^2}{2}} \, {\bf \Lambda}} \;,  
\label{sol-UV}
\end{eqnarray} 
where an arbitrary and unphysical overall phase has been fixed by demanding that ${\hat X}_{\rm in} $ is real and positive at some given (rescaled) early time $x_{\rm in}$ during the adiabatic regime.

\begin{figure}[tbp]
\centering 
\includegraphics[width=0.6\textwidth,angle=0]{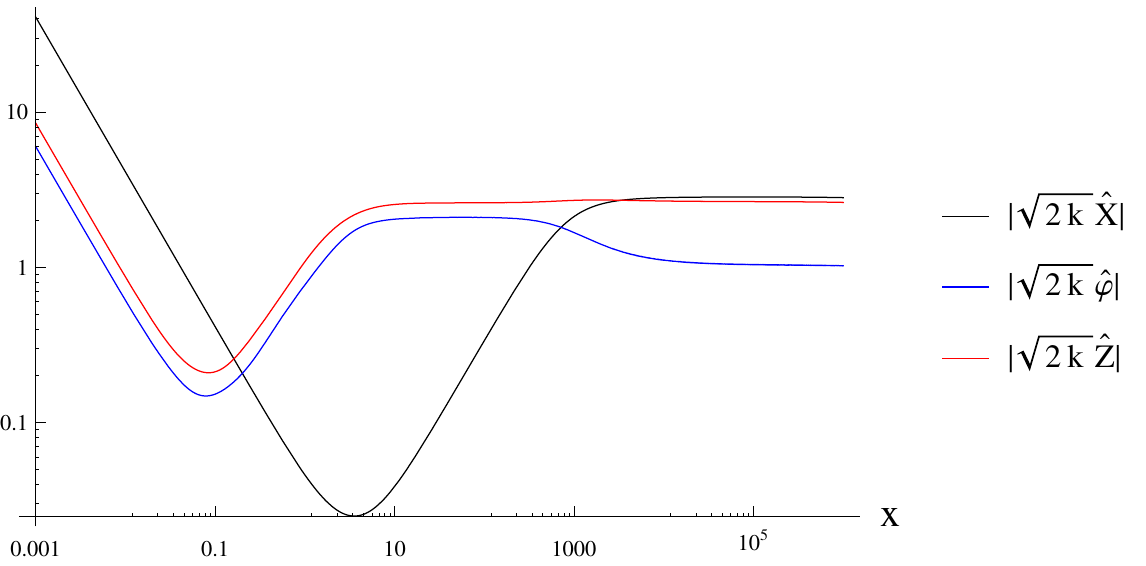}
\caption{
Evolution of the scalar perturbations (\ref{scalar-canonical}) obtained from the linearized equations (\ref{4.11-4.12-4.13-us}), and with initial conditions (\ref{sol-UV}). The parameters in the evolutions are ${\tilde f} = 0.1 ,\, \lambda = 1000 ,\, \frac{g}{{\tilde \mu}^2} = 20$. This gives ${\bf \Lambda } \simeq 340$ and $m_Q \simeq 2.56$ at $N=60$ e-folds of inflation.} 
\label{fig:scalar-lin}
\end{figure}

The linear solutions of equations (\ref{4.11-4.12-4.13-us}), with initial conditions according to (\ref{sol-UV}) are shown in Figure \ref{fig:scalar-lin} for a given choice of parameters. We recall that time flows from right to left in this plot (since $x \equiv - k \tau$). We note that the adiabatic solution (with nearly constant amplitudes) persists until $x \ga {\bf \Lambda}$. We also see that ${\hat X}$ scales as $a \simeq \frac{1}{x}$ ouside the horizon, resulting in a constant inflaton perturbation $\delta \chi$ in this regime.

%%%%%%%%%%%%%%%%%%%%%%%%%%%%%%%%%%%%%
\subsection{Phenomenology from the linearized computation} 
\label{subsec:pheno} 
%%%%%%%%%%%%%%%%%%%%%%%%%%%%%%%%%%%%%

As done in \cite{Dimastrogiovanni:2012ew,Adshead:2013nka} we assume that, after inflation, only the inflaton field provides a sizable contribution to reheating (we note that the energy in the gauge field is much smaller than the inflaton energy during inflation). In this limit, we have the curvature perturbation (in the spatially flat gauge that we are using) 
\begin{equation}
\zeta = - \frac{H \, \delta \chi}{\dot{\chi}} \simeq \frac{\lambda}{f} \, \frac{H \tau {\hat X}}{2} \, \frac{m_Q}{1+m_Q^2}  \;. 
\end{equation} 

This results in the linear power spectrum
\begin{equation}
P_\zeta^{(0)} \equiv \frac{k^3}{2 \pi^2} \vert \zeta \vert^2 = 
\frac{1+\cos {\tilde x}_N}{48 \pi^2} \, \frac{\lambda^2 \, {\tilde \mu}^4}{{\tilde f}^2} \, \frac{m_Q^2}{\left( 1 + m_Q^2 \right)^2} \, 
\left\vert x \; \sqrt{2 k} {\hat X} \right\vert^2  \;. 
\label{P-zeta-lin} 
\end{equation} 
Imposing that the power spectrum at the scales that leave the horizon at $N=60$ is equal to the measured value, $P_\zeta \simeq 2.2 \cdot 10^{-9}$ \cite{Ade:2015lrj} provides the value of ${\tilde \mu}$. 

We can then compute the tensor-to-scalar ratio $r_{\rm linear} \equiv \frac{P_L+P_R}{P_\zeta}$ and the special tilt $n_{s,{\rm linear}} \equiv 1 + \frac{d \ln P_\zeta}{d \ln k} \simeq 1 - \frac{d \ln P_\zeta}{d N} \,$. (In the last expression we evaluate $k=aH$ at horizon crossing, and we disregard the variation of $H$ with respect to that of $a$.) For simplicity, we evaluate them both at the same scale, corresponding to $N=60$.  The former is obtained by dividing the GW power provided in eqs. (\ref{PL}) and (\ref{PR}) by the measured value of the scalar power spectrum. The latter is obtained by two contributions. One in which we differentiate (using the slow roll expressions) the background-dependent quantities entering in (\ref{P-zeta-lin}). One in which we take the differential between the value of ${\hat X}$ obtain numerically at $N=60$ and at $N=59$.

\begin{figure}[tbp]
\centering 
\includegraphics[width=0.6\textwidth,angle=0]{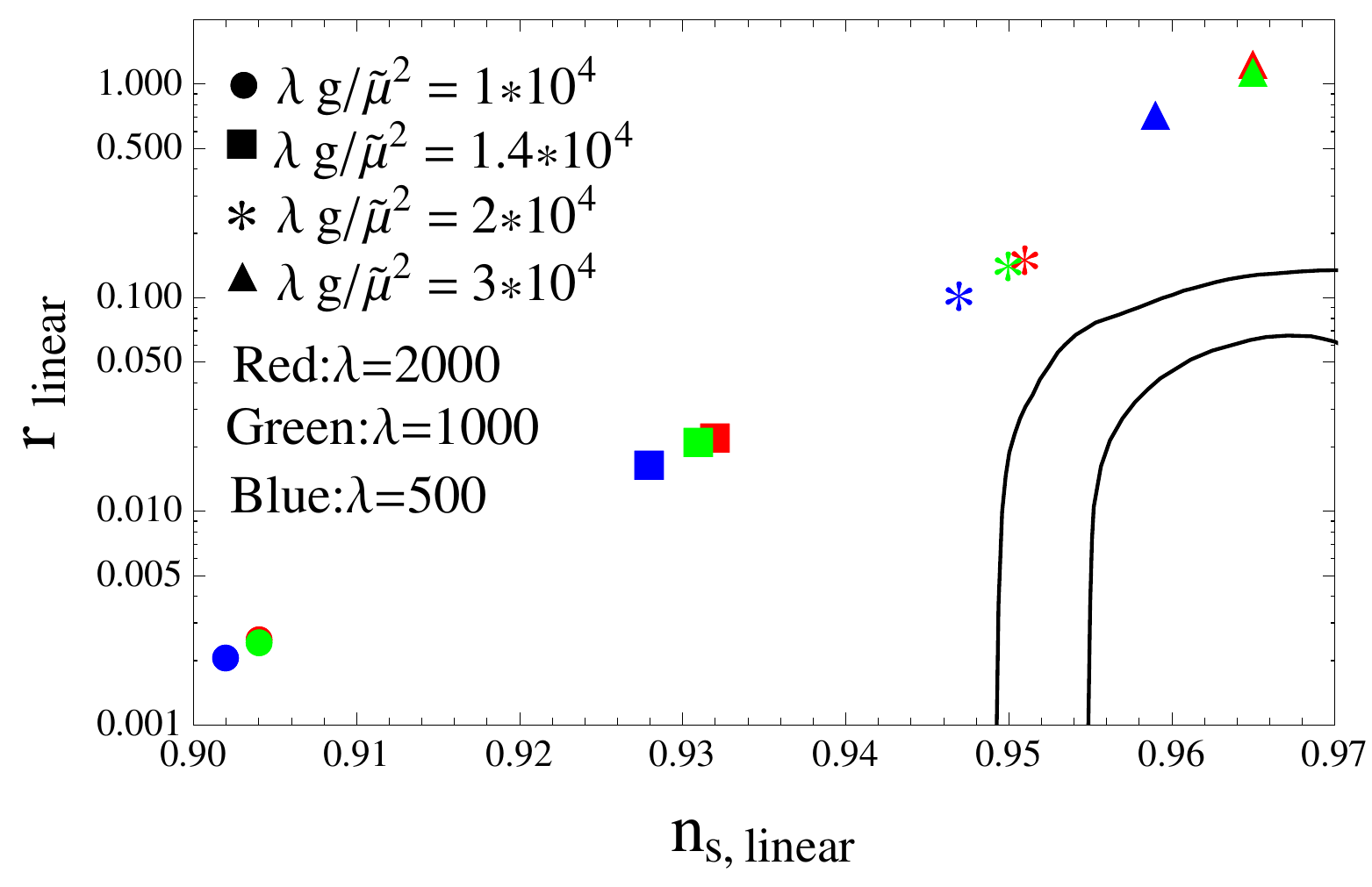}
\caption{$\left\{ n_{s,{\rm linear}} - r_{\rm linear} \right\}$ obtained for ${\tilde f} = 0.1$ and for the choice of parameters indicated in the figure. Similar values are obtained at fixed $\lambda g / {\tilde \mu}^2$. The linear tensor-to-scalar ratio is obtained from the linearized theory (disregarding the nonlinear contribution to the scalar perturbations discussed in this work). All points are outside the Planck contours \cite{Ade:2015lrj},  also shown in the plot.  All cases shown assume $N=60$ e-folds of inflation. } 
\label{fig:ns-r}
\end{figure}

The results are shown in Figure \ref{fig:ns-r}, for ${\tilde f} = 0.1$, and for the choice of parameters reported in the Figure. As already remarked in the Introduction, the results from the linearized theory are  incompatible with observations \cite{Dimastrogiovanni:2012ew,Adshead:2013qp,Adshead:2013nka}.

%%%%%%%%%%%%%%%%%%%%%%%%%%%%%%%%%%%%%
\section{One nonlinear interaction} 
\label{sec:nonlinear} 
%%%%%%%%%%%%%%%%%%%%%%%%%%%%%%%%%%%%%

In this section we single out one specific nonlinear interaction, namely the one between the enhanced tensor mode $t_L$ of the SU(2) multiplet, and 
the inflaton perturbation. This interaction originates from the terms $\propto F^a F^a$ and $\propto  \frac{\lambda}{f} \, \chi F^a {\tilde F}^a$ in the Lagrangian. 
As remarked in the Introduction, we conjecture that these are the dominant nonlinear interaction since they have a coupling proportional to~\footnote{For the  $F^a F^a$ term, the coefficient   $\frac{\lambda}{f} $ is due to the dependence of $\delta A_0^a$ on $\chi$, see eq. (\ref{deco-interaction}).}  $\frac{\lambda}{f}$, which is much greater (since $\lambda \gg 1$, and $f \ll M_p$) than the gravitational interactions, since the mode $t_L$ is also the origin of the GW enhancement, and since the inflaton perturbation is eventually the one that is mapped into the observed scalar perturbation. 

The enhanced tensor mode $t_L$ also interacts with the two other dynamical modes ${\hat Z}$ and ${\hat \varphi}$.  It is hard to imagine that the inclusion of additional interactions would provide a modification of the power spectrum that precisely cancel against the one found here, and it is more reasonable to expect that additional interactions could lead to a further departure from the linear theory considered in the previous section. In this light, our result might be considered as a lower bound on the modification of the linear theory. 

To extract the interaction of our interest, we decompose 
\begin{equation}
\chi = \chi \left( t \right) + \delta \chi \;\;,\;\; 
A_0^a \supset  - \frac{\lambda}{f} \frac{g a^2 Q^2}{- \partial^2+2 g^2 a^2 Q^2} \partial_a \delta \chi \equiv  \partial_a {\tilde \chi}  \;\;,\;\; A_i^a \supset \delta_i^a \, a \left( t \right) \,  Q \left( t \right) + t_{ia}  \;, 
\label{deco-interaction} 
\end{equation}
where $t_{ia}$ is symmetric, transverse, and traceless. 

Due to the fact that $A_0^a$ contains a term proportional to $\delta \chi$, and due to the non-abelian nature of the gauge field,  the vector kinetic term induces the following interactions 
\begin{equation}
\sqrt{-g} \, {\cal L} \supset - \frac{\sqrt{-g}}{4} F_{\mu \nu}^a F^{\mu \nu,a} \supset  g \, \epsilon^{abc} t'_{ia} t_{ic} \, \partial_b  {\tilde \chi} + {\rm O } \left( \delta \chi^2 \, t \right)  + {\rm O } \left( \delta \chi^2 \, t^2 \right) \;.  
\label{interaction1}
\end{equation} 
The first term contributes to the one-loop diagram shown in Figure \ref{fig:diagrams}. The ${\rm O } \left( \delta \chi^2 \, t \right)  $ term gives rise to another one-loop diagram, with one $\delta \chi$ line and one $t$ line as propagators. 
This second diagram is suppressed with respect to the one shown in Figure \ref{fig:diagrams} since it has one fewer $t$ propagator, and therefore it is proportional to two fewer powers of the mode function $t_L$ (the sourced term studied in this work becomes significant precisely due to the exponential enhancement of $t_L$ with $m_Q$, see Figure \ref{fig:tmax}). 
The third term gives rise to another one-loop diagram with a single $t$ propagator, with both ends attached to the same point  on the $\delta \chi^2$ line. This diagram is also suppressed by two fewer powers of the mode function $t_L$.~\footnote{ The second term can also be combined with a $t^3$ interaction to give a tadpole diagram, made by one $t$ propagator with one end on the $\delta \chi^2$ line, and the other end on a $t$ loop. This propagator has zero momentum, and the sourced mode $t_L$ is not enhanced in this limit.} 

Other ${\rm O } \left( \delta \chi \, t^2 \right)$ interactions are obtained from the $\phi F {\tilde F}$ term. We note that we can rewrite this term as 
\begin{equation}
 -  \frac{\lambda}{8 f} \chi \epsilon^{\mu \nu \rho \sigma} F_{\mu \nu}^a F_{\rho \sigma}^a =   \frac{\lambda}{f} \chi \partial_\sigma J^\sigma \;\;,\;\; 
J^\sigma = \epsilon^{\mu \nu \rho \sigma} \left( \frac{1}{2} A_\mu^a \partial_\nu A_\rho^a  + \frac{g}{6} \epsilon_{abc} A_\mu^a A_\nu^b A_\rho^c \right) \;. 
\end{equation} 
When we expand this term, and evaluate all the  ${\rm O } \left( \delta \chi \, t^2 \right)$ contributions, we find that the term  $\propto \delta \chi$ contained in  $A_0^a $ can be collected in a total derivative, and can therefore be disregarded. Ignoring these terms, one finds 
\begin{equation}
J_0 \supset - \frac{1}{2} \epsilon^{ijk} t_{ai} \partial_j t_{ak}  + \frac{g a Q }{2} \, t_{ab}  t_{ab} \;\;,\;\; 
J_i \supset - \frac{1}{2} \epsilon^{ijk} t_{aj} t_{ak}' \;, 
\end{equation} 
and, combining the last two equations, 
\begin{equation}
 - \frac{\lambda}{8 f} \chi \epsilon^{\mu \nu \rho \sigma} F_{\mu \nu}^a F_{\rho \sigma}^a \supset \frac{\lambda}{f} \, \delta \chi  \; \left[ \frac{g}{2}  \left( a Q \, t_{ab} \, t_{ab} \right)'  -  \epsilon^{ijk} \, t_{ai}' \,  \partial_j t_{ak} \right]  \;. 
\label{interaction2}
\end{equation} 

The right hand sides of eqs. (\ref{interaction1}) and  (\ref{interaction2}) contain all and only all the terms that are linear in $\delta \chi$ and quadratic in $t_{ai}$. These terms  define our interaction hamiltonian 

\begin{equation}
H_{\rm int} = - \frac{\lambda}{f} \int d^3 x \left\{ \delta \chi \left[ \frac{g}{2} \left( a Q t_{ab} t_{ab} \right)' 
- \epsilon^{ijk} t_{ai}' \partial_j t_{ak} \right] + \left[  \frac{g^2 a^2 Q^2}{- \partial^2+2 g^2 a^2 Q^2} \delta \chi \right] \, \partial_j \left( \epsilon^{ijk} t_{ia}' t_{ak} \right)  \right\} \;, 
\label{Hint}
\end{equation}
that we employ to compute the corrections to the inflaton correlation function through 
\begin{eqnarray} 
&& \delta \left\langle \delta \chi \left( \tau ,\, \vec{k}_1 \right) \,  \delta \chi \left( \tau ,\, \vec{k}_2 \right) \right\rangle 
= - \int^\tau d \tau_1 \, \int^{\tau_1} d \tau_2 \nonumber\\ 
&& \quad\quad\quad\quad 
\times \left\langle \left[ \left[  \delta \chi^{(0)} \left( \tau ,\, \vec{k}_1 \right) \,  \delta \chi^{(0)} \left( \tau ,\, \vec{k}_2 \right) 
,\, H_{\rm int}^{(0)} \left( \tau_1 \right) \right] ,\,  H_{\rm int}^{(0)} \left( \tau_2 \right) \right] \right\rangle  \;. 
\label{in-in} 
\end{eqnarray} 
The suffix $(0)$ remarks that the mode functions entering at the right hand side are the ``unperturbed'' ones, namely those obtained in the linear theory presented in the previous section. For brevity, we will omit this suffix from now on. It is convenient to express the unperturbed modes in terms of the dimensionless mode functions $X_c$ and $t_c$, defined through 
\begin{eqnarray} 
\delta \chi \left( \tau ,\, k \right) &\equiv& \frac{\sqrt{1+m_Q^2}}{\sqrt{2 k}} \, \frac{X_c \left( x \right)}{a \left( \tau \right)} = 
\sqrt{\frac{1+m_Q^2}{2}} \, \frac{H}{k^{3/2}} \, x \, X_c \left( x \right) \;, \nonumber\\ 
{\hat t}^+ \left( \tau ,\, k \right) &\equiv& \frac{t_c \left( x \right)}{\sqrt{2 k}} \;\;\;,\;\;\; x \equiv - k \tau \; . 
\label{code-var} 
\end{eqnarray} 
They correspond to the canonically normalized variables, times $\sqrt{2 k}$. Therefore, their initial amplitude is $1$, and they are function of the dimensionless quantity $x \equiv - k \tau$.

In Appendix \ref{app:B} we show that 
\begin{equation}
 \delta \left\langle \delta \chi \left( \tau ,\, \vec{k}_1 \right) \,  \delta \chi \left( \tau ,\, \vec{k}_2 \right) \right\rangle' 
= \frac{\lambda^2}{f^2}  \int^\tau d \tau_1 \, \int^{\tau_1} d \tau_2 
\int \frac{d^3 p_1 d^3 p_2}{\left( 2 \pi \right)^3} \, 
\frac{\left( {\hat p}_1 \cdot {\hat p}_2 - 1 \right)^4}{16} \, \delta^{(3)} \left( \vec{k}_1 - \vec{p}_1 - \vec{p}_2 \right)  \, {\cal T} \;,     
\label{dP-chi} 
\end{equation}

where the prime on the left hand side denotes the correlator without the corresponding $\delta \left( \vec{k_1} + \vec{k}_2 \right)$ function, and where ${\cal T}$ is the real function 
\begin{eqnarray} 
{\cal T} &\equiv& 
\frac{H^4 \left( 1 + m_Q^2 \right)^2 x^2}{8 k_1^4 q_1 q_2} \, {\rm Re } 
\left[ X_c \left( x \right) X_c^* \left( x_2 \right) \left( X_c \left( x_1 \right) X_c^* \left( x \right) - {\rm c.c.} \right) {\cal W} \left( x_1 ,\, x_2 ,\, q_1 ,\, q_2 \right) \right] \;. \nonumber\\ 
\label{calT}  
\end{eqnarray} 
In this expression, we have defined 
\begin{eqnarray} 
&&   \!\!\!\!\!\!\!\!  \!\!\!\!\!\!\!\! 
{\cal W} \left( x_1 ,\, x_2 ,\, q_1 ,\, q_2 \right) \equiv 
 \frac{m_Q^2}{x_1 \, x_2}  \, 
t_c \left( q_2 x_1 \right) t_c^* \left( q_2 x_2 \right) t_c \left( q_1 x_1 \right) t_c^* \left( q_1 x_2 \right) \nonumber\\ 
&&   \!\!\!\!\!\!\!\! 
+  2  \left[ m_Q -x_1\, q_1  + \left( q_1-q_2 \right) x_1 \, \frac{m_Q^2}{x_1^2 + 2 m_Q^2} \right]   \, q_2 \,  t_c^{\;'} \left( q_2 \, x_1 \right) t_c^* \left( q_2 \, x_2 \right)  \,  t_c \left( q_1 \, x_1 \right) \nonumber\\
&& \times \left\{ - \frac{m_Q}{x_2} \, t_c^* \left( q_1 \, x_2 \right) + q_1 \left[ m_Q - x_2 \, q_2   + \left( q_2-q_1 \right) x_2 \frac{m_Q^2}{x_2^2 + 2 m_Q^2} \right]  t_c^{\;'*} \left( q_1 \, x_2 \right) \right\}  \nonumber\\ 
 &&   \!\!\!\!\!\!\!\! 
+ 2   \left[ m_Q - x_2 \, q_2  + \left( q_2-q_1 \right) x_2 \frac{m_Q^2}{x_2^2 + 2 m_Q^2} \right]  \, q_1 \, t_c \left( q_2 \, x_1 \right) \, t_c^* \left( q_2 \, x_2 \right) \,  t_c^{\;'*} \left( q_1 \, x_2 \right) \nonumber\\
&&
 \times  \left\{ - \frac{m_Q}{x_1} \, t_c \left( q_1 \, x_1 \right) + q_1 \left[ m_Q - x_1 \, q_2  + \left( q_2-q_1 \right) x_1 \frac{m_Q^2}{x_1^2 + 2 m_Q^2} \right]  t_c^{\;'} \left( q_1 \, x_1 \right) \right\}  \;, 
\label{calW}  
\end{eqnarray} 
where $q_i \equiv \frac{p_i}{k_1}  ,\; x_i \equiv - k_1 \tau_i $.

We are interested in the ratio between this nonlinear contribution to the power spectrum and the linear term: 
\begin{equation}
{\cal R}_{\delta \chi} \equiv \frac{\delta P_\chi \left( \tau ,\, k \right)}{P_\chi \left( \tau ,\, k \right)} = \frac{ \delta \left\langle \delta \chi \left( \tau ,\, \vec{k}_1 \right) \,  \delta \chi \left( \tau ,\, \vec{k}_2 \right) \right\rangle' }{\left\langle \delta \chi \left( \tau ,\, \vec{k}_1 \right) \,  \delta \chi \left( \tau ,\, \vec{k}_2 \right) \right\rangle' }  \;, 
\label{Rdchi-def}
\end{equation} 
Working out the expression (\ref{dP-chi})  at the numerator  (see  Appendix \ref{app:B}) gives 
\begin{eqnarray} 
 \!\!\!\!\!\!\!\! R_{\delta \chi}   &  = &
\frac{\lambda^2 \, H^2}{256 \pi^2  f^2} \,   \frac{1+m_Q^2}{\left\vert  X_c \left( x \right) \right\vert^2} \, 
\int_x d x_1 \, \int_{x_1} d x_2  \int_0^\infty d q_1  \, \int_{\vert 1 - q_1 \vert}^{1 + q_1} d q_2  \, \left( \frac{1 - \left( q_1 + q_2 \right)^2}{2 q_1 q_2} \right)^4 \nonumber\\ 
& & \times  \, {\rm Re } 
\left[ X_c \left( x \right) X_c^* \left( x_2 \right) \left( X_c \left( x_1 \right) X_c^* \left( x \right) - {\rm c.c.} \right) {\cal W} \left( x_1 ,\, x_2 ,\, q_1 ,\, q_2 \right) \right] \;. 
\label{R-chi}
\end{eqnarray} 
We note that this expression is manifestly dimensionless and real. Moreover, since $X_c \left( x \right) \propto a \propto \frac{1}{x}$ outside the horizon, and since an equal number of $X_c \left( x \right)$ are contained at the numerator and at the denominator, this expression is also constant in time well outside the horizon (this is the case provided that the integrand of the $\int d x_1$ integration is not peaked in the IR. We verified that this is the case; in practice, the nonlinear source vanishes when it is well super-horizon). As we discuss below, this ratio is scale-dependent, due to the evolution of $m_Q$ during inflation.~\footnote{We compute the scalar perturbations $\delta \chi$ produced by the enhanced ${\hat t}^+$ modes. Therefore, to consistently describe this effect, we exclude from the integration domain the times for which the enhancement has yet to take place (specifically, the values of $x$ that are so large so that the term in parenthesis in eq. (\ref{eq-t}) is still positive, and the bump visible in Figure \ref{fig:tmax} has yet to take place).  This provides an upper bound on the arguments $q_i x_j$ of the modes entering in ${\cal W}$ and, ultimately, on the $x_1$ and $x_2$ integrations. We verified that changing the precise location of the upper bound impacts the final result at a negligible level, since most of the support of the integral occurs at times for which  the mode functions ${\hat t}^+$ have been well enhanced by the tachyonic growth, and not at the start of this growth. } 

The first line is symmetric under the exchange $q_1 \leftrightarrow q_2$, so we symmetrize also the second line. Moreover, we can disentangle the extrema of integration of the residual spatial integration by defining the combinations 
\begin{equation}
{\cal X} \equiv \frac{q_1+q_2}{\sqrt{2}} \;\;,\;\;   {\cal Y}   \equiv \frac{q_1-q_2}{\sqrt{2}} \;. 
\end{equation}
In terms of these variables, 
\begin{eqnarray} 
 \!\!\!\!\!\!\!\! R_{\delta \chi}   &  = &
\frac{\lambda^2 \, H^2}{256 \pi^2  f^2} \,   \frac{1+m_Q^2}{\left\vert  X_c \left( x \right) \right\vert^2} \, 
\int_x d x_1 \, \int_{x_1} d x_2  \int_{\frac{1}{\sqrt{2}}}^\infty d {\cal X}  \, \int_0^{\frac{1}{\sqrt{2}}} d {\cal Y}  \, \left( \frac{1 - 2 {\cal X}^2}{{\cal X}^2 - {\cal Y}^2} \right)^4  \, {\rm Re } \Bigg\{ X_c \left( x \right) X_c^* \left( x_2 \right) \nonumber\\
& & \quad\quad \times  \left[ X_c \left( x_1 \right) X_c^* \left( x \right) - {\rm c.c.} \right] 
\left[ {\cal W} \left( x_1 ,\, x_2 ,\, \frac{{\cal X}+{\cal Y}}{\sqrt{2}} ,\, \frac{ {\cal X}-{\cal Y}}{\sqrt{2}} \right) +  {\cal W} \left( x_1 ,\, x_2 ,\, \frac{{\cal X}-{\cal Y}}{\sqrt{2}} ,\, \frac{{\cal X}+{\cal Y}}{\sqrt{2}} \right) \right] \Bigg\} \;. \nonumber\\ 
\end{eqnarray}

\begin{figure}[tbp]
\centering 
\includegraphics[width=0.45\textwidth,angle=0]{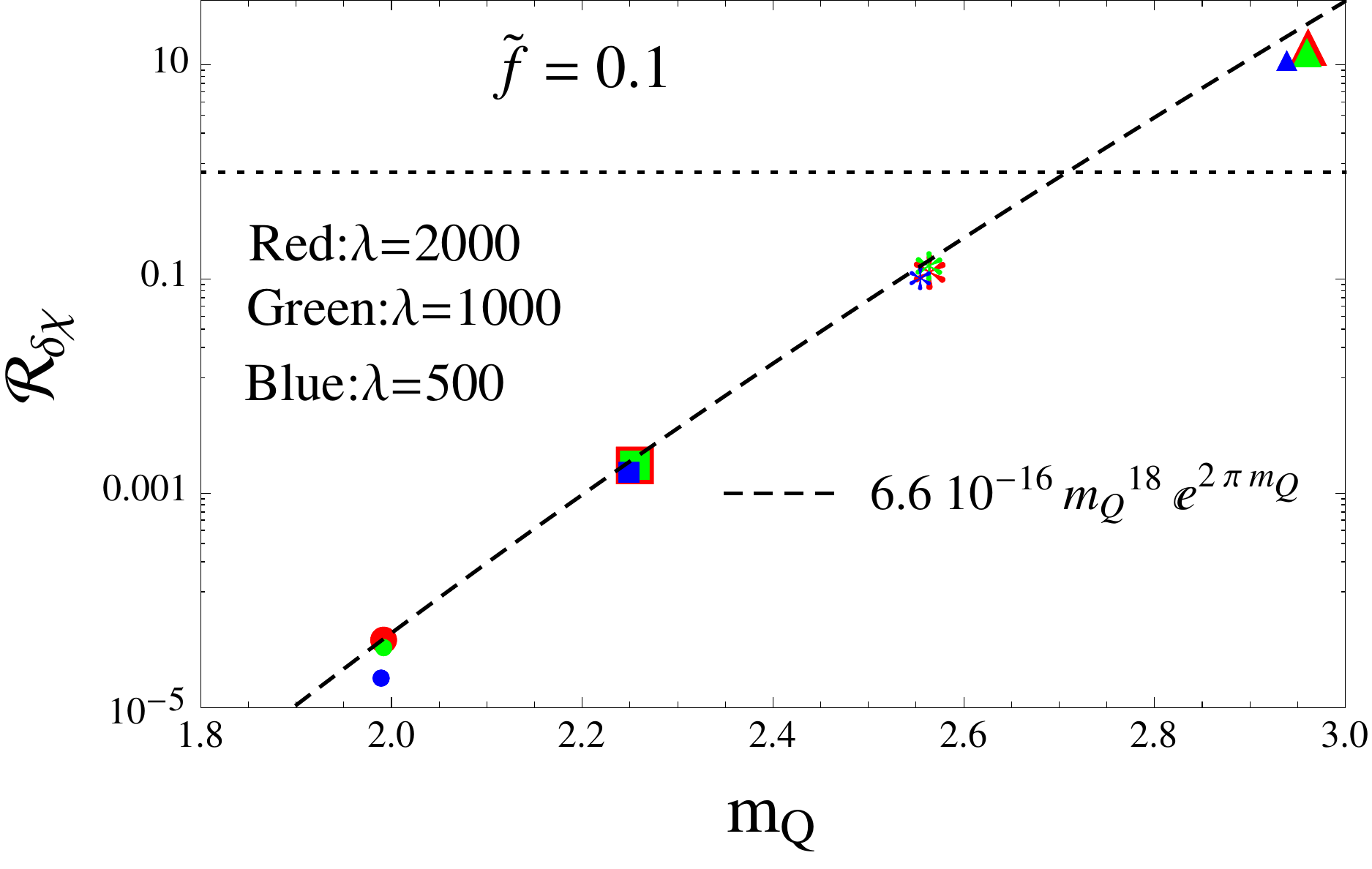}
\includegraphics[width=0.45\textwidth,angle=0]{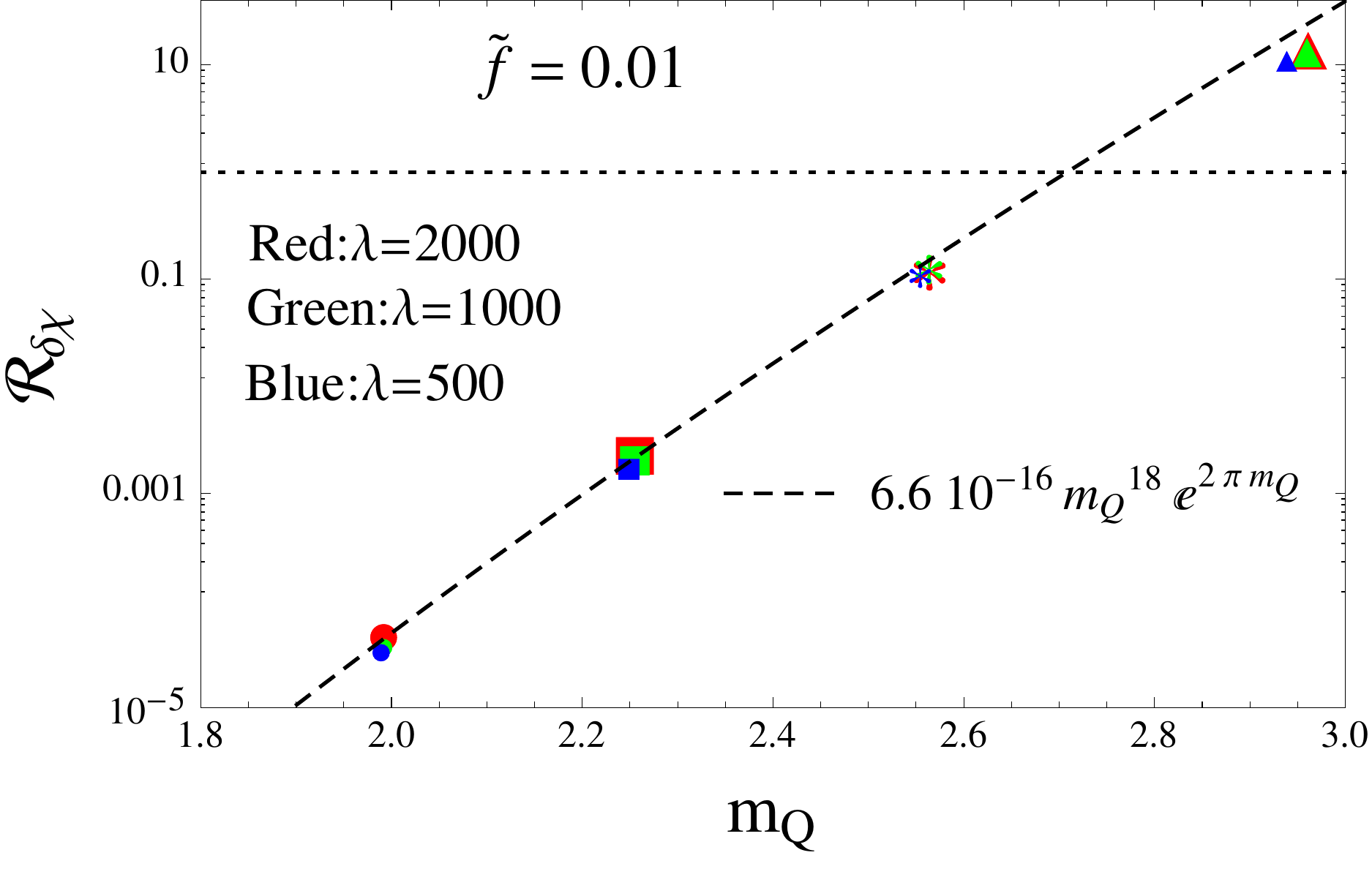}
\caption{Ratio between the nonlinear correction to the power spectrum studied in this section and the linear power spectrum (eq. (\ref{Rdchi-def})) as a function of $m_Q$, for the same choice of parameters as in Figure  \ref{fig:ns-r}. The two panels differ for the choice of ${\tilde f}$. Different choices of parameters that result in the same $m_Q$ give nearly the same $R_{\delta \chi}$. The results are well reproduced by the analytic relation (\ref{Rdchi-analytic}), shown as a dashed line in the figure. 
We stress that the portion of this figure with ${\cal R}_{\delta \chi} \ga 1$ is inconsistent, since the parameter ${\tilde \mu}$ 
has been fixed by assuming that the linear perturbations dominate at CMB scales. 
} 
\label{fig:mQ-R}
\end{figure}

The result is shown in Figure \ref{fig:mQ-R} for different choices of parameters, and for $x=10^{-2}$ (the ratio becomes constant at super-horizon scales). We note that $R_{\delta \chi}$ is mostly sensitive to the value of $m_Q$, and that  different choices of parameters that result in the same $m_Q$ give nearly the same $R_{\delta \chi}$. In Appendix \ref{sec:appC} we provide a semi-analytic study of the results presented in the figure, obtaining 
\begin{equation}
R_{\delta \chi}    \simeq  6.6 \cdot 10^{-16} \, m_Q^{18} \, {\rm e}^{2 \pi m_Q} \;,  
\label{Rdchi-analytic}
\end{equation} 
in the range $1.8 \leq m_Q \leq 3$ that we have studied in this work. We note that  this relation, and the results presented in Figure \ref{fig:mQ-R}, are only valid if $R_{\delta \chi} < 1$ at CMB scales, since we assumed that the vacuum modes dominate the power spectrum in setting the value of ${\tilde \mu}$ from the power spectrum normalization. 

The exponential dependence on $m_Q$ of eq. (\ref{Rdchi-analytic}) is due to the fact that (i) the maximum amplitude of the amplified tensor mode scales as ${\rm e}^{\pi m_Q /2}$ (see eq. (\ref{tmaxapprox}) and Figure \ref{fig:tmax}), and that (ii) $\delta P_\chi$
is proportional to four powers of the tensor mode (given that $\delta \chi$ is sourced by two modes $t$). 
An analogous dependence is obtained in the U(1) case \cite{Barnaby:2010vf} (where the parameter $\xi$ plays an analogous role to $m_Q$ of this model). The $m_Q^{18}$ dependence emerges from a series of factors, including the fact that several combination of the parameters of the model can be written in terms of $m_Q$ alone along the inflationary solution, and that the central position and amplitude of the bump in the tensor mode, and the amplitude of the scalar mode scale with $m_Q$. We discuss this in details in Appendix \ref{sec:appC}. 

The approximate relation (\ref{Rdchi-analytic}) reproduces very well the numerical results, as can be seen from the figure. 

As we mentioned, our result is valid as long as $R_{\delta \chi} $ is significantly smaller than one on CMB scales. Given the strong dependence on the result on $m_Q$, a small change of $m_Q$ is enough to change from the $R_{\delta \chi} \la 1$ to the $R_{\delta \chi} \ll 1$ regime. Therefore, we simply set $R_{\delta \chi} \la 1$ , which implies $m_Q \la 2.7$.  Combining this with the result (\ref{mQ-N}), we see that 
\begin{equation}
R_{\delta \chi} \la 1  \;\;\;\; \longleftrightarrow \;\;\;\;  m_Q \la 2.7  \;\;\;\; \longleftrightarrow \;\;\;\; 
\frac{\lambda g}{{\tilde \mu}^2} \la 2 \cdot 10^4 \, \left( \frac{N}{60} \right)^{3/2} \;, 
\end{equation} 
where we recall that $N$ is the number of e-folds before the end of inflation at which the modes for which $R_{\delta \chi}$ is computed left the horizon. 

We note from eq. (\ref{mQ-N}) that $m_Q$ increases during inflation, 
\begin{equation}
m_Q \left( N \right) = m_{Q,{\rm CMB}} \, \sqrt{\frac{N_{\rm CMB}}{N}} \;, 
\label{mQ-N-2}
\end{equation}
so that the sourced perturbations have a blue spectrum, and can come to dominate over the vacuum ones at small scales, even if they are subdominant at CMB scales. We discuss the consequences of this scale dependence in the reminder of this section. 

\subsection{Scale dependence of the sourced perturbations, and phenomenological implications} 

We conclude this section by commenting on the scale dependence of the total power spectrum. Parametrizing the linear power spectrum in terms of amplitude $P_{\rm CMB}$, linear spectral tilt $n_{sL}$, and running of the linear spectral tilt $\alpha_{sL}$ at the CMB scales, and using the relations (\ref{Rdchi-analytic}) and (\ref{mQ-N-2})  for the nonlinear contribution to the power spectrum, we have 
\begin{eqnarray} 
P_{\zeta,{\rm tot}} \left( k_N \right) &\simeq& P_{\rm CMB} \, {\rm e}^{\left( 60 - N \right) \left( n_{s,L} - 1 \right) + \frac{\left( 60 - N \right)^2}{2} \, \alpha_{s,L}} \, \left[ 1 + {\cal R}_{\delta \chi} \left( N \right) \right] \;\;, \nonumber\\ 
{\cal R}_{\delta \chi} \left( N \right) &\simeq & 
6.6 \cdot 10^{-16} \left[ m_{Q,{\rm CMB}} \, \sqrt{\frac{60}{N}} \right]^{18}  \, {\rm e}^{2 \pi \, m_{Q,{\rm CMB}} \, \sqrt{\frac{60}{N}} } \, , 
\label{Pzeta-tot}
\end{eqnarray} 
where we recall that $N$ is the number of e-folds before the end of inflation at which the mode $k_N$ exited the horizon. 
In parametrizing the linear power spectrum, we have disregarded the variation of $H$ with respect to that of the scale factor, and  we have assumed $N_{\rm CMB} = 60$. 

The spectral tilt and running of the total power spectrum (\ref{Pzeta-tot}), evaluated at CMB scales, are then given by 
\begin{eqnarray} 
n_s &\simeq& n_{s,L} + \left[ - \frac{{\cal R}'_{\delta \chi} }{1+{\cal R}_{\delta \chi} }  \right]_{N = 60} \equiv n_{s,L} +  n_{s,NL} \;, \nonumber\\ 
\alpha_s &\simeq&  \alpha_{s,L} 
+ \left[  \frac{{\cal R}''_{\delta \chi} }{1+{\cal R}_{\delta \chi} } 
- \left(  \frac{{\cal R}'_{\delta \chi} }{1+{\cal R}_{\delta \chi} } \right)^2 
\right]_{N=60}  \equiv \alpha_{s,L} +  \alpha_{s,NL} \;. 
\label{ns-as-tot}
\end{eqnarray}

\begin{figure}[tbp]
\centering 
\includegraphics[width=0.48\textwidth,angle=0]{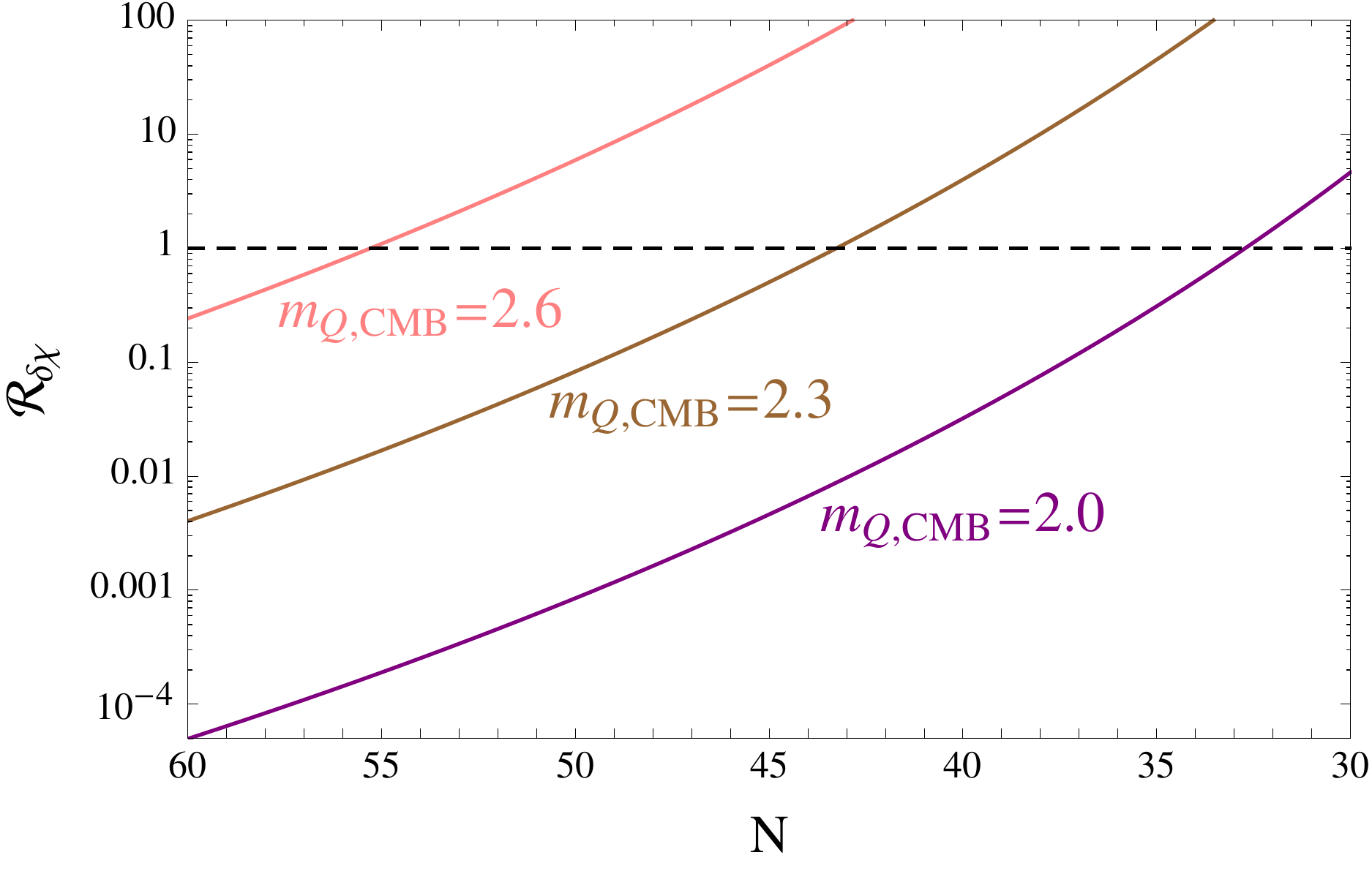}
\includegraphics[width=0.45\textwidth,angle=0]{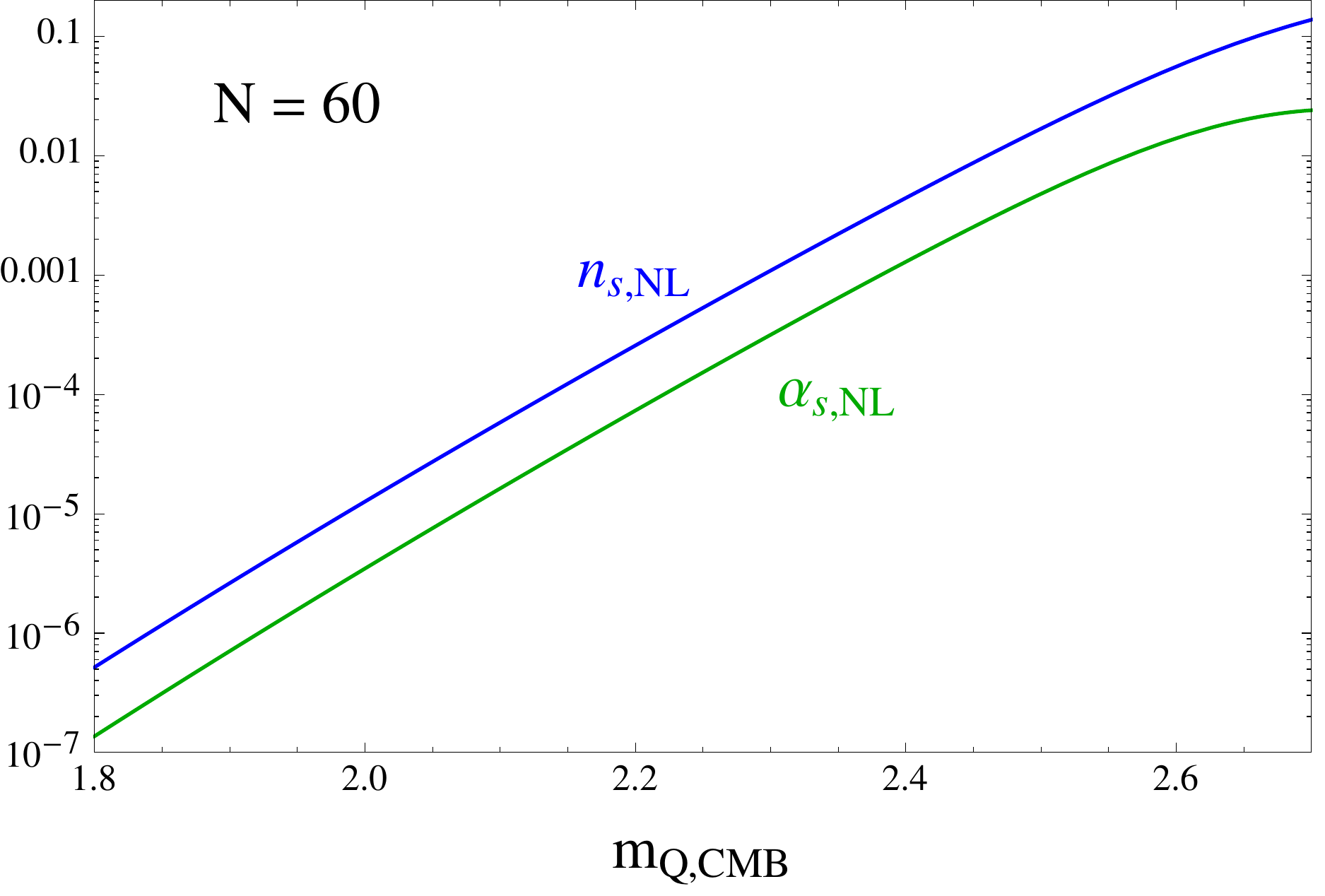}
\caption{Left panel: scale dependence of the ratio between the nonlinear correction to the power spectrum studied in this section and linear power spectrum (eq. (\ref{Rdchi-def})).The horizontal axis denotes the number of e-folds before the end of inflation at which one given mode exits the horizon. The three lines shown correspond to different choices of parameters, that result in three different values of $m_Q$ at CMB scales. Notice that the range of scales shown by this figure is much greater than just those measured in the CMB. Right panel: Effect of the growth for CMB experiments. Specifically, we show the nonlinear contributions to the scalar spectral tilt and its running, evaluated at CMB scales $(N=60)$. 
} 
\label{fig:Rchi-N}
\end{figure}

In the left panel of Figure \ref{fig:Rchi-N} we show the scale dependence of ${\cal R}_{\delta \chi}$ for three choices of parameters for which the nonlinear contribution to the power spectrum that we have computed is subdominant at CMB scales ($N = 60$). The horizontal axis denotes the number of e-folds $N$ at which one given mode leaves the horizon, and the vertical axis denotes the value of ${\cal R}_{\delta \chi}$ for that mode. In the right panel we show the contribution to the scalar spectral tilt and running from this nonlinear term, evaluated at CMB scales (using  the terms in square bracket in eqs. (\ref{ns-as-tot})).

\begin{figure}[tbp]
\centering 
\includegraphics[width=0.6\textwidth,angle=0]{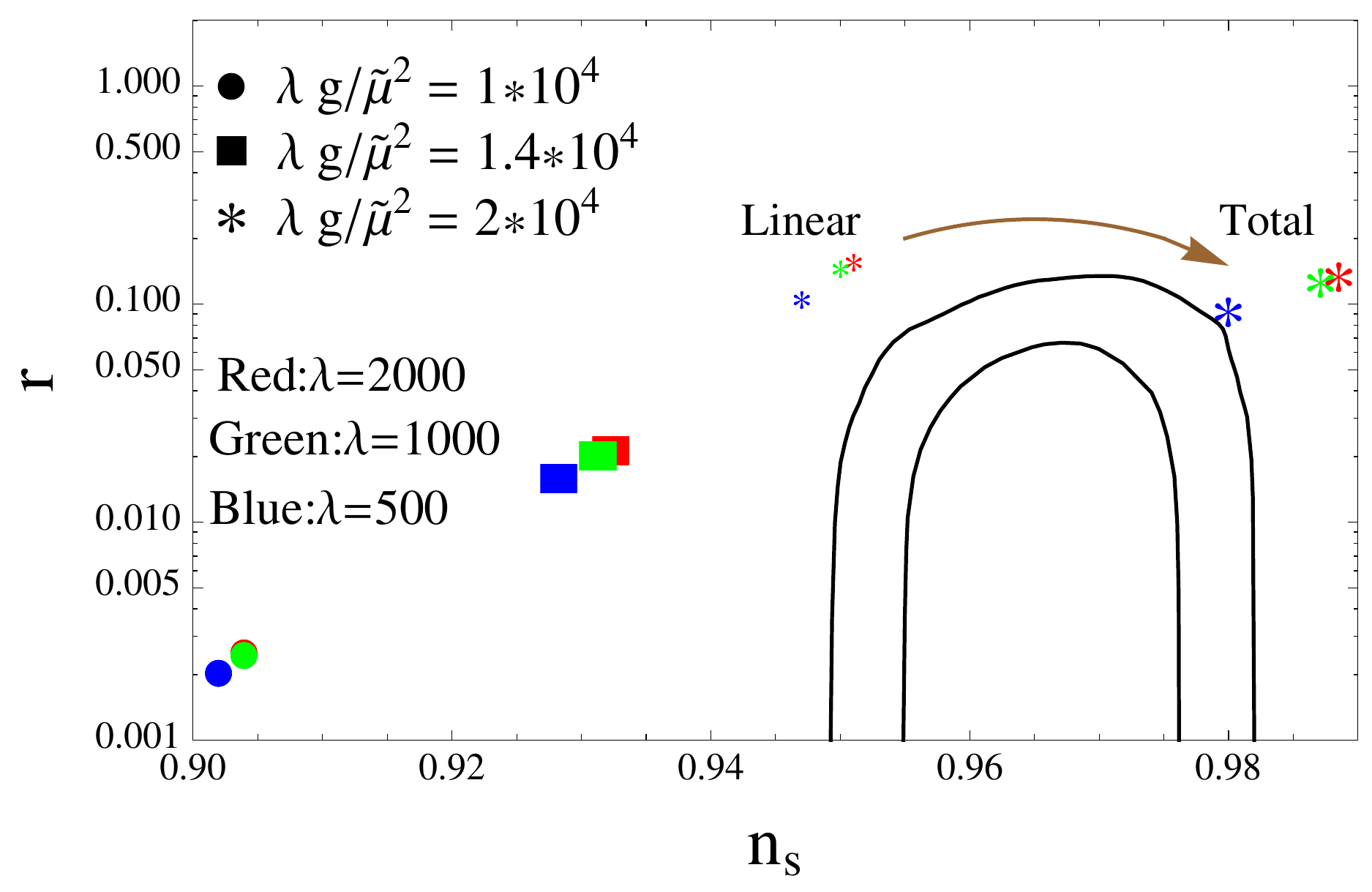}
\caption{$\left\{ n_s - r_{\rm linear} \right\}$ obtained for the first three set of points already shown in 
Figure \ref{fig:ns-r}. As compared to that figure, we now also show the effect of the nonlinear interaction studied in this section. For the first two sets of points (indicated by circles and squares) the nonlinear contribution is negligible, and 
the result is superimposed to the one of the linear theory. For the third set (indicated by asterisks) the effect is significant, and it is indicated by the arrow; the point in smaller (respectively, larger) size correspond to the results of the linear (respectively, linear plus nonlinear) computation. } 
\label{fig:ns-r-tot}
\end{figure}

We see from Figure  \ref{fig:ns-r} that the parameter choices that result in an allowed value for the tensor-to-scalar ratio $r$ provide a too red spectrum (too small $n_s$). On the other hand,  we see from Figure \ref{fig:Rchi-N} that the nonlinear term is blue. It is natural to ask wether its addition can reconcile the model with the data. The nonlinear contributions to $n_s$ and to $r = \frac{P_{\rm GW}}{P_{\zeta,{rm tot}}}$ dramatically increase with $m_Q$, so this effect switches from negligible to very relevant within a small variation of $m_Q$. We show this in Figure \ref{fig:ns-r-tot}. The first two sets of points shown with circles and squares are characterized, respectively, by $m_Q \simeq 1.99$ and $m_Q \simeq 2.25$. For these values the nonlinear contributions are negligible, and the results shown in the figure practically coincide with those from the linear theory already shown in Figure  \ref{fig:ns-r}. The third set of points shown with asterisks is instead characterized by $m_Q \simeq 2.56$. For this value, the nonlinear contribution to the spectral tilt is significant, $n_{s,{\rm NL}} \simeq 0.035$. This causes the change indicated by the arrow in the figure. This improves the agreement with the data, giving rise to a small portion of parameters that would be within the $2 \sigma$ Planck contours. For such value of $m_Q$, the nonlinear contribution to the power spectrum that we have computed amounts to about $14\%$ of the linear power. Based on the results from the U(1) models, we expect that the nonlinear contribution is highly non-Gaussian  \cite{Barnaby:2010vf}. Therefore, such a large ratio might be excluded from the stringent limits on non-Gaussianity. This computation is beyond the scope of this work. 

Proceeding to smaller scales than CMB, the large growth of the nonlinear contribution shown in the left panel of Figure 
\ref{fig:Rchi-N} can have an impact on Large Scale Structure, CMB $\mu-$ and $y-$ distortions, and primordial black holes (in analogy with the growth of the sourced signal obtained in the U(1) case, where the effects have been studied in several works). Such considerations should be discussed in the context of models that are in better agreement with the CMB than the original Chromo-Natural Inflation model studied in this work.

%%%%%%%%%%%%%%%%%%%%%%%%%%%%%%%%%%%%%
\section{Conclusions}
\label{sec:conclusions} 
%%%%%%%%%%%%%%%%%%%%%%%%%%%%%%%%%%%%%

In this work we have performed the first step toward the computation of nonlinear effects involving scalar perturbations in models where the inflaton is coupled to a non-Abelian vector field carrying a nonvanishing spatial vev. As a prototype, we focused our attention on the original model of Chromo-Natural Inflation. However, we expect that this computation could be extended with minor modifications also to the more recent implementations of this mechanism, constructed to improve the agreement with the CMB data. For simplicity, we restricted our computations to the interaction between the tensor mode $t_L$, that is amplified at horizon crossing, and the scalar perturbation $\delta \chi$ (the perturbation of the inflaton), that coincides with the adiabatic scalar perturbation at large-scales. A full computation requires the inclusion of also the two other scalar modes of the model (that originate from linear combinations of the SU(2) fields); generically, we do not expect that different nonlinear diagrams will cancel against each other, so we argue that our result should be considered as a lower-bound estimate for the full nonlinear correction. 

The most important property of the nonlinear correction that we computed is its strong scale-dependence. The sourcing mode $t_L$ is exponentially sensitive to the parameter $m_Q$ (defined in eq. (\ref{mqdef})),  which coincides,  in the large $m_Q$ regime, with the parameter $\xi \equiv \frac{\lambda \dot{\chi}}{2 f H}$ that characterizes the nonlinear effects in the U(1) case. This parameter grows during inflation, so that the nonlinear contribution strongly grows at smaller scales, see Figure \ref{fig:Rchi-N}. For the specific case of Chromo-Natural Inflation, this growth can by beneficial for the comparison against the CMB data. From the linearized studies, the model is ruled out by the CMB comparison, since the scalar spectrum is too red in the range of parameters that produce a sufficiently small amount of GW. Including the blue contribution that we have computed can bring the predictions of the model inside the Planck $2 \sigma$ contour, for a narrow choice of parameters. In this regime, the nonlinear correction amounts to ${\rm O } \left( 10 \% \right)$ of the 
linear term. Based on the results from the U(1) models, we expect  the nonlinear contribution to be highly non-Gaussian. An order $10\%$ completely non-Gaussian contribution (which is likely peaked at equilateral configurations, as in the Abelian case  \cite{Barnaby:2010vf}) is probably incompatible with the CMB data, so that this study is in order before claiming a better overall agreement with the CMB. 

Moreover, the strong scale dependence of the nonlinear term results in a large running of the spectral tilt. The largest value of $m_Q$ considered in the points shown in Figure \ref{fig:ns-r-tot}, results in a contribution to the spectral tilt from the nonlinear term of $n_{s,{\rm NL}} \simeq 0.035$, and to the running of $\alpha_{s,{\rm NL}} \simeq 0.0095 $. In generic models of slow-roll inflation the running appears only at second order in slow roll, and it is therefore expected to be of ${\rm O } \left( \left( n_s - 1\right)^2 \right)$.  In the U(1) case, CMB limits from the grow of the power spectrum provide a comparable, or possibly even stronger constraint than that from CMB non-Gaussianity \cite{Meerburg:2012id,Ade:2015lrj}. Going to progressively smaller scales, this growth can have implications for Large Scale Structure, CMB $\mu-$ and $y-$ distortions, and primordial black holes. 

For any given model, studying the limits from the scalar sector is crucial for a consistent discussion of the GW phenomenology. This has been well established in the U(1) mechanisms, where the limits from the scalar sector generically prevent the sourced GW to be observed at CMB scale \cite{Barnaby:2012xt,Ferreira:2014zia}, with the exception of only very few special constructions \cite{Namba:2015gja,Fujita:2017jwq}. Once a more complete analysis of the nonlinearities in the scalar sector has been obtained, such limits should also be studied for the various modifications of Chromo-Natural Inflation emerged in the recent literature. 

To summarize, we have performed a first step in the computation of nonlinear scalar perturbations in models where an axion inflaton is coupled to a non-Abelian multiplet of gauge fields with nonvanishing spatial vev. We computed the variation of the power spectrum of Chromo-Natural Inflation resulting from the coupling of the inflaton to the enhanced tensor mode $t_L$. Following analogous studies performed in the U(1) case, this result can be extended in a number of directions by: performing a full analysis, that includes all scalar perturbations; studying different models in this class; extending this computation to the three-point function; finally, considering the various phenomenological implications of this signal (which is typically exponentially sensitive to the growing inflaton speed). We hope to come back to some of these points in future works. 

\vskip.25cm

{\bf Note added}: As the computations presented in this work were completed, and the present manuscript was being written, ref. \cite{Dimastrogiovanni:2018xnn} appeared on the archive. This work computes the scalar-tensor-tensor bispectrum arising at tree level from the coupling $t_L \, t_L \, \delta \chi$, where $\delta \chi$ is an axion field different from the inflaton.

\vskip.25cm
\section*{Acknowledgements} 

We thank Lorenzo Sorbo for very useful discussions. The work of M.P. is partially supported from the DOE grant DE-SC0011842  at the University of Minnesota.

\vskip.25cm

\appendix

%%%%%%%%%%%%%%%%%%%%
\section{Initial conditions for the linear scalar perturbations }
\label{app:A}
%%%%%%%%%%%%%%%%%%%%

In this Appendix we derived the early time adiabatic linearized scalar solution (\ref{sol-UV}). In the UV ($x \rightarrow \infty$) regime,  the system (\ref{4.11-4.12-4.13-us})  reduces to 
\begin{equation} 
\left\{ \begin{array}{l} 
{\hat X}'' + {\hat X} + \frac{\sqrt{2} {\bf \Lambda}}{x} {\hat \varphi} + \frac{\sqrt{2} {\bf \Lambda} m_Q}{x} \, {\hat Z}' \simeq 0 \;, \\ 
{\hat \varphi}'' + {\hat \varphi} +  \frac{\sqrt{2} {\bf \Lambda}}{x} {\hat X} \simeq 0 \;, \\ 
{\hat Z}'' + {\hat Z} -  \frac{\sqrt{2} {\bf \Lambda} m_Q}{x} {\hat X}' \simeq 0 \;. 
\end{array} \right. 
\label{eq-WKB}
\end{equation} 
We note that  a term $\propto {\hat Z}$ has not been included in the second equation (although having the same  $\frac{1}{x}$ parametric dependence as another mass term that we have retained), since it is not enhanced by the large parameter ${\bf \Lambda} \gg 1$. 

To solve (\ref{eq-WKB}),  we look for (WKB) solutions of the type 
\begin{equation}
{\hat X} = A \left( x \right) \, {\rm e}^{i \kappa \left( x \right)} \;,\; 
{\hat \varphi} = B \left( x \right) \, {\rm e}^{i \kappa \left( x \right)} \;,\; 
{\hat z} = C \left( x \right) \, {\rm e}^{i \kappa \left( x \right)}  \;, 
\end{equation} 
and we neglect $\kappa'' ,\, A' ,\, B' ,\, C'$ (to be justified below). This gives 
\begin{eqnarray}
&& A+\frac{\sqrt{2}\Lambda}{x}B+\frac{i \sqrt{2}m_Q \Lambda \kappa'(x)}{x}C- {\kappa'(x)}^2 A = 0 \;, \nonumber\\
&& B+\frac{\sqrt{2}\Lambda}{x}A-{\kappa'(x)}^2 B = 0 \;, \nonumber\\ 
&& C-\frac{i\sqrt{2}m_Q \Lambda \kappa'(x)}{x}A-{\kappa'(x)}^2 C = 0 \;. 
\label{eq-WKB2} 
\end{eqnarray}

We rewrite this as a matrix equation for the $\left( \begin{array}{c} A \\ B \\ C \end{array} \right)$ vector. Imposing that the determinant of the matrix vanishes gives 
\begin{equation}
\kappa' = 1 \;\;,\;\; 
\kappa'  = \left\{ 1 + \frac{m_Q^2 {\bf \Lambda}^2}{x^2}  \pm \frac{{\bf \Lambda} \sqrt{2 \left( m_Q^2+1 \right)  + \frac{m_Q^4 {\bf \Lambda}^2}{x^2}}}{x} \right\}^{1/2} \;. 
\end{equation} 
Following  \cite{Adshead:2013nka}, we disregard the first solution, as it enforces $A=0$ (no inflaton perturbation). We also disregard the fast oscillating mode (the one with the plus sign in the curly bracket) as it matches to a decaying mode with amplitude $\propto \sqrt{x}$. 

For the slow oscillating mode, the equations (\ref{eq-WKB2}) are solved by 
\begin{eqnarray}
B &=& \frac{\sqrt{2} {\bf \Lambda}}{x \left( \kappa^{'2} - 1 \right)} \, A =  - A \left[ \frac{1}{\sqrt{1+m_Q^2}} + \frac{m_Q^2 {\bf \Lambda}}{\sqrt{2} \, \left( 1 + m_Q^2 \right) x} + {\rm O } \left( \frac{1}{x^2} \right) \right] \;\;, \nonumber\\ 
C &=& - \frac{i \sqrt{2} {\bf \Lambda} \, \kappa'}{x \left( \kappa^{'2} - 1 \right)} \, A =   i A \left[ \frac{m_Q}{\sqrt{1+m_Q^2}} - \frac{m_Q {\bf \Lambda}}{\sqrt{2} \, \left( 1 + m_Q^2 \right) x} + {\rm O } \left( \frac{1}{x^2} \right) \right] \;. 
\end{eqnarray} 

We also Taylor expand $\kappa' = 1 - \frac{{\bf \Lambda} \sqrt{1+m_Q^2}}{\sqrt{2} x} + {\rm O } \left( \frac{1}{x^2} \right)$, and integrate it from a reference early time $x_{\rm in}$ to $x$, to obtain 
\begin{equation}
\kappa = x - x_{\rm in} + \sqrt{\frac{1+m_Q^2}{2}} \, {\bf \Lambda} \, \ln \left( \frac{x_{\rm in}}{x} \right) \;. 
\end{equation} 
(One can then verify that the terms we have suppressed in (\ref{eq-WKB2}) are of ${\rm O } \left( \frac{1}{x^2} \right)$, which is the same order that we have disregarded in writing (\ref{eq-WKB}).) Altogether, we find 
\begin{eqnarray}
{\hat X} &=& A \,  {\rm e}^{i \left( x - x_{\rm in} \right)} \, \left( \frac{x_{\rm in}}{x} \right)^{i\sqrt{ \frac{1+m_Q^2}{2}} \, {\bf \Lambda}} \;, \nonumber\\ 
{\hat \varphi} &=& - \frac{A}{\sqrt{1+m_Q^2}} \,   {\rm e}^{i \left( x - x_{\rm in} \right)} \, \left( \frac{x_{\rm in}}{x} \right)^{i \sqrt{\frac{1+m_Q^2}{2}} \, {\bf \Lambda}} \;, \nonumber\\ 
{\hat z} &=&  \frac{i A m_Q}{\sqrt{1+m_Q^2}} \,   {\rm e}^{i \left( x - x_{\rm in} \right)} \, \left( \frac{x_{\rm in}}{x} \right)^{i \sqrt{\frac{1+m_Q^2}{2}} \, {\bf \Lambda}} \;, 
\end{eqnarray} 
in agreement with \cite{Adshead:2013nka}. 

We then note that the the system (\ref{eq-WKB}) could have been derived by an early time effective action, in which the mode ${\hat \varphi}$ has standard canonical conjugate momentum, $\Pi_{\hat \varphi} = {\hat \varphi}'$. This is not the case for the other two modes. 
This leads us to impose that this mode has the standard adiabatic normalization, and therefore $A = \frac{\sqrt{1+m_Q^2}}{\sqrt{2 k}}$. This leads to the early time adiabatic solution  (\ref{sol-UV}) reported in the main text.

%%%%%%%%%%%%%%%%%%%%
\section{Evaluation of $\delta \left\langle \delta \chi^2 \right\rangle$ through the in-in formalism }
\label{app:B}
%%%%%%%%%%%%%%%%%%%%

In this Appendix we present the derivation of eq. (\ref{dP-chi}), starting from eqs.  (\ref{Hint}) and (\ref{in-in}). 
We decompose the inflaton and the tensor perturbations as 
\begin{eqnarray} 
\delta {\hat \chi} \left( \tau ,\, \vec{k} \right) &=& \delta \chi \left( \tau , k \right) a_\chi \left( \vec{k} \right) +  \delta \chi_k^* \left( \tau , k \right) a_\chi^\dagger  \left( - \vec{k} \right) \;, \nonumber\\ 
{\hat t}_{ab}  \left( \tau ,\, \vec{k} \right) &=& \sum_{\lambda=\pm} \Pi_{ab,\lambda}^* \left( {\hat k} \right) 
\left[ {\hat t}^\lambda \left( \tau ,\, k \right) a_\lambda \left( \vec{k} \right) +  {\hat t}^{\lambda*} \left( \tau ,\, k \right) a_\lambda^\dagger \left( - \vec{k} \right) \right] 
\equiv  \sum_{\lambda=\pm} \Pi_{ab,\lambda}^* \left( {\hat k} \right) \, {\hat {\hat t}}_\lambda \left( t ,\, \vec{k} \right) 
 \;. \nonumber\\ 
\label{chi-t-deco}
\end{eqnarray} 
The operators at the left hand side are the Fourier transform (\ref{FT}) of the fields entering at the right hand side of (\ref{in-in}). 
The annihilation / creation operators satisfy the nonvanishing relations $\left[ a_\chi \left( \vec{k} \right) ,\,  a_\chi^\dagger \left( \vec{k}'  \right) \right] =\delta^{(3)} \left( \vec{k} - \vec{k}' \right)$ and $\left[ a_\lambda \left( \vec{k} \right) ,\,  a_{\lambda'}^\dagger \left( \vec{k}'  \right) \right] =\delta^{(3)} \left( \vec{k} - \vec{k}' \right) \delta_{\lambda \lambda'}$.  The mode function of the inflaton is related to the canonical variable ${\hat X}$ defined in eq. (\ref{scalar-canonical}) by $\delta \chi = \frac{{\hat X}}{a}$. The sum in the second line is over the left ($\lambda = +$) and right ($\lambda = -$) handed helicities.  The transverse, traceless, and symmetric tensor operators can be written as 
\begin{equation}
{\Pi_{ab,\lambda}}^*  \left( {\hat k} \right) \equiv \epsilon_{a,\lambda} \left( {\hat k} \right)  \epsilon_{b,\lambda} \left( {\hat k} \right) \;, 
\end{equation} 
where the  vector circular polarization operators satisfy $\vec{k} \cdot \vec{\epsilon}_\lambda \left( {\hat k} \right)= 0 ,\, \vec{k} \times \vec{\epsilon}_\lambda \left( {\hat k} \right) = - i \lambda k \vec{\epsilon}_\lambda \left( {\hat k } \right) $, $ \vec{\epsilon}_\lambda \left( - {\hat k } \right) =  \vec{\epsilon}_\lambda^{\;*} \left(  {\hat k } \right) ,\,  \vec{\epsilon}_\lambda^{\;*} \left( {\hat k } \right) \cdot  \vec{\epsilon}_{\lambda'} \left( {\hat k } \right) = \delta_{\lambda \lambda'} $, in addition to $\vec{\epsilon}_+ \left( {\hat k} \right) \cdot \vec{\epsilon}_+ \left( {\hat k} \right) = \vec{\epsilon}_- \left( {\hat k} \right) \cdot \vec{\epsilon}_- \left( {\hat k} \right) = 0$. Thanks to these properties, the mode functions ${\hat t}^{\pm}$ are canonically normalized, and coincide with those introduced in eq. (\ref{tensor-canonical}).~\footnote{The simplest way to see this is to note that, for a wave-vector oriented in the third direction, $\vec{k} = \left( 0 ,\, 0 ,\, k \right)$, the operators $\vec{\epsilon}_\pm = \left( \frac{1}{\sqrt{2}} ,\, \pm \frac{i}{\sqrt{2}} ,\, 0 \right)$ satisfy all the properties we just wrote. This then enforces $\Pi_\pm^* \left( {\hat k} \right) = \frac{1}{2} \left( \begin{array}{ccc} 1 &  \pm i & 0 \\  \pm i & - 1 & 0 \\ 0 & 0 & 0 \end{array} \right)$, and the mode functions obtained from (\ref{chi-t-deco}) coincide with those written in (\ref{tensor-modes}). } In the following, we only consider the enhanced ${\hat t}^+$ mode, and therefore keep only the term $\lambda = +$ in eq. (\ref{chi-t-deco}). 

Keeping all this into account, the interaction Hamiltonian can be rewritten as 
\begin{eqnarray}
H_{\rm int} \left( \tau \right) & = & - \frac{\lambda}{2 f} \int \frac{d^3 p_1 d^3 p_2}{\left( 2 \pi \right)^{3/2}} \left[ \vec{\epsilon}_+ \left( {\hat p}_1 \right) \cdot 
 \vec{\epsilon}_+ \left( {\hat p}_2 \right) \right]^2 
\bigg\{ \left[ g \, a \, Q - p_2 +\left(p_2-p_1 \right) \frac{g^2 a^2 Q^2}{\vert {\vec p_1} + {\vec p_2} \vert^2 + 2 g^2 a^2 Q^2 } \right] \partial_{\tau'} +  \nonumber\\  
& & \!\!\!\!\!\!\!\!\!\!\!\!\!\!\!\!\!\!\!\!\!\!\!\!\!\!\!\!\!\!\!\!\!\!\!\!\!\!\!\! \left[ g \, a \, Q - p_1  + \left(p_1-p_2 \right) \frac{g^2 a^2 Q^2}{\vert {\vec p_1} + {\vec p_2} \vert^2 + 2 g^2 a^2 Q^2 } \right] \partial_{\tau''} + g \left[  a \, Q \right]'  \bigg\}  \, \delta {\hat \chi} \left( \tau ,\, - \vec{p}_1 -   \vec{p}_2 \right) \, {\hat {\hat t}}^+ \left( \tau' ,\, \vec{p}_1 \right) 
\, {\hat {\hat t}}^+ \left( \tau'' ,\, \vec{p}_2 \right) \Bigg\vert_{\tau' = \tau'' = \tau} \;. 
\end{eqnarray}

From this Hamiltonian, the correction (\ref{in-in}) to the inflaton $2-$point function acquires the form 
\begin{eqnarray}
&& \delta \left\langle \delta \chi \left( \tau ,\, \vec{k}_1 \right) \,  \delta \chi \left( \tau ,\, \vec{k}_2 \right) \right\rangle 
=   \frac{\lambda^2}{4 f^2} \int^\tau d \tau_1 \, \int^{\tau_1} d \tau_2 \nonumber\\ 
&& \quad\quad \times \int \frac{d^3 p_1 d^3 p_2}{\left( 2 \pi \right)^{3/2}} \,  \left[ \vec{\epsilon}_+ \left( {\hat p}_1 \right) \cdot  \vec{\epsilon}_+ \left( {\hat p}_2 \right) \right]^2 \; 
\bigg\{ \left[ g \, a  \, Q \left( \tau_1 \right)  - p_2   + \left(p_2-p_1 \right) \frac{g^2 a^2 Q^2}{\vert {\vec p_1} + {\vec p_2} \vert^2 + 2 g^2 a^2 Q^2 }  \right] \partial_{\tau_1'}  
\nonumber\\ 
&& \quad\quad +  \left[ g \, a  \, Q \left( \tau_1 \right)  - p_1  +  \left(p_1-p_2 \right) \frac{g^2 a^2 Q^2}{\vert {\vec p_1} + {\vec p_2} \vert^2 + 2 g^2 a^2 Q^2 }  \right] \partial_{\tau_1''} + g \, \partial_{\tau_1} \left[ a  \, Q \left( \tau_1 \right) \right] \bigg\} \nonumber\\
&& \quad\quad \times 
\int \frac{d^3 p_3 d^3 p_4}{\left( 2 \pi \right)^{3/2}} \,  \left[ \vec{\epsilon}_+ \left( {\hat p}_3 \right) \cdot  \vec{\epsilon}_+ \left( {\hat p}_4 \right) \right]^2 \; 
\bigg\{ \left[ g \, a \, Q \left( \tau_2 \right)  - p_3  +  \left(p_3-p_4 \right) \frac{g^2 a^2 Q^2}{\vert {\vec p_3} + {\vec p_4} \vert^2 + 2 g^2 a^2 Q^2}  \right] \partial_{\tau_2'}   \nonumber\\ 
&& \quad\quad +  \left[ g \, a \, Q \left( \tau_2 \right)  - p_4  +  \left(p_4-p_3 \right) \frac{g^2 a^2 Q^2}{\vert {\vec p_3} + {\vec p_4} \vert^2 + 2 g^2 a^2 Q^2}  \right] \partial_{\tau_2''} + g \, \partial_{\tau_2} \left[ a \, Q \left( \tau_2 \right) \right] \bigg\} \nonumber\\
&& \quad\quad \times \Bigg\{ \left[ {\cal C}_\chi \left( \tau_1 ,\, \tau ;\, k_1 \right) - {\rm c.c.} \right] \delta^{(3)} \left( \vec{k}_1 - \vec{p}_1 - \vec{p}_2 \right) \,  \delta^{(3)}\left( \vec{k}_2 - \vec{p}_3 - \vec{p}_4 \right) 
\Bigg[  {\cal C}_\chi \left( \tau ,\, \tau_2 ;\, k_2 \right) \nonumber\\ 
&& \quad\quad\quad\quad \times \Bigg(  {\cal C}_+\left( \tau'_1 ,\, \tau''_2 \, ; p_1\right)\, {\cal C}_+\left( \tau''_1 ,\, \tau'_2 \, ; p_2\right)\delta^{(3)} \left( \vec{p}_1 + \vec{p}_4\right)\delta^{(3)} \left( \vec{p}_2 + \vec{p}_3 \right)\nonumber\\
&&    \quad\quad\quad\quad\quad + {\cal C}_+\left( \tau'_1 ,\, \tau'_2 \, ; p_1\right)\, {\cal C}_+\left( \tau''_1 ,\, \tau''_2 \, ; p_2\right)\delta^{(3)} \left( \vec{p}_1 + \vec{p}_3\right)\delta^{(3)} \left( \vec{p}_2 + \vec{p}_4\right) \Bigg) - {\rm c.c.} \Bigg]  \nonumber\\
&&    \quad\quad\quad\quad      + \left( \vec{k}_1 \, \leftrightarrow \vec{k}_2 \right) \Bigg\} \;\;\;\; \Bigg\vert_{\tau_1' = \tau_1'' = \tau_1 \;,\;\; \tau_2' = \tau_2'' = \tau_2 } \;, 
\label{deltaP-par} 
\end{eqnarray} 
where we defined 
\begin{equation}
\left\langle  \delta \chi^{(0)} \left( \tau ,\, \vec{k} \right)   \delta \chi^{(0)} \left( \tau' ,\, \vec{k}' \right) \right\rangle = \delta \chi \left( \tau ,\, k \right) \delta \chi^* \left( \tau' ,\, k \right) \delta^{(3)} \left( \vec{k} + \vec{k}' \right) \equiv {\cal C}_\chi \left( \tau ,\, \tau' ;\, k \right) \,  \delta^{(3)} \left( \vec{k} + \vec{k}' \right) \;, 
\end{equation} 
and analogously for ${\cal C}_+$ in terms of the mode functions ${\hat t}^+$. 

Using the fact that the third and fourth lines of (\ref{deltaP-par}) are invariant under the simultaneous $\vec{p}_3 \leftrightarrow \vec{p}_4$ and $\tau_2' \leftrightarrow \tau_2''$, the fifth and sixth line give the same contribution to the result. The $\int d^3 p_3 d^3 p_4$ integrations then produce 
\begin{eqnarray}
&& \delta \left\langle \delta \chi \left( \tau ,\, \vec{k}_1 \right) \,  \delta \chi \left( \tau ,\, \vec{k}_2 \right) \right\rangle 
=   \frac{\lambda^2}{2 f^2} \delta^{(3)} \left( \vec{k}_1 + \vec{k}_2 \right)  \int^\tau d \tau_1 \, \int^{\tau_1} d \tau_2 
\int \frac{d^3 p_1 d^3 p_2}{\left( 2 \pi \right)^3} \,  \left\vert \vec{\epsilon}_+ \left( {\hat p}_1 \right) \cdot  \vec{\epsilon}_+ \left( {\hat p}_2 \right) \right\vert^4  \nonumber\\ 
& & \quad\quad \quad\quad\quad\quad  \quad\quad\quad\quad  \quad\quad\quad\quad  \quad\quad\quad\quad 
 \left[ \delta^{(3)} \left( \vec{k}_1 - \vec{p}_1 - \vec{p}_2 \right) +  \delta^{(3)} \left( \vec{k}_2 - \vec{p}_1 - \vec{p}_2 \right) \right] 
\; {\cal T} \;, 
\label{deltaP-par2} 
\end{eqnarray} 
where
\begin{eqnarray} 
{\cal T} &\equiv& 2 \, {\rm Re } \Bigg[ \left\{ {\cal C}_\chi \left( \tau_1 ,\, \tau ; k_1 \right) - {\rm c.c.} \right\} {\cal C}_\chi \left( \tau ,\, \tau_2 ; k_1 \right) \nonumber\\ 
& & \quad \times \Bigg\{ 
g^2 \, \partial_{\tau_1} \left[ a \, Q \left( \tau_1 \right)  \right] \, \partial_{\tau_2} \left[ a \, Q \left( \tau_2 \right) \right] \, {\cal C}_+ \left( \tau_1 ,\, \tau_2 ;\, p_2 \right) \, {\cal C}_+ \left( \tau_1 ,\, \tau_2 ;\, p_1 \right) \nonumber\\ 
&&  \!\!\!\!\!\!\!\!\!\!\!\!\!\!\!\!\!\!\!\!\!\!\!\!\!\!\!\!\!\!\!\!\!\!\!\!\!\!\!\!\!\!\! \!\!\!\!\!\!\!\! 
+  2 \, \left[ g \, a \, Q \left( \tau_1 \right) - p_1   +  \left(p_1-p_2 \right) \frac{g^2 a^2 Q^2(\tau_1)}{\vert {\vec p_1} + {\vec p_2} \vert^2 + 2 g^2 a^2 Q^2 \left( \tau_1 \right)} \right] \, \partial_{\tau_1} {\cal C}_+ \left( \tau_1 ,\, \tau_2 ;\, p_2 \right)
 \nonumber\\ 
 &&  \!\!\!\!\!\!\!\!\!\!\!\!\!\!\!\!\!\!\!\!\!\!\!\!\!\!\!\!\!\!\!\!\!\!  \!\!\!\!\!\!\!\! \times  \, \left\{ \left[ g \, a \, Q \left( \tau_2 \right) - p_2   + \left(p_2-p_1 \right) \frac{g^2 a^2 Q^2(\tau_2)}{\vert {\vec p_1} + {\vec p_2} \vert^2 + 2 g^2 a^2 Q^2 \left( \tau_2 \right)}  \right] \, \partial_{\tau_2}\,  {\cal C}_+ \left( \tau_1 ,\, \tau_2 ;\, p_1 \right) + \partial_{\tau_2}\, \left[ g \, a \, Q \left( \tau_2 \right)  \right] \,  {\cal C}_+ \left( \tau_1 ,\, \tau_2 ;\, p_1 \right) \right\} \nonumber\\
 &&  \!\!\!\!\!\!\!\!\!\!\!\!\!\!\!\!\!\!\!\!\!\!\!\!\!\!\!\!\!\!\!\!\!\!\!\!\!\!\!\!\!\!\! \!\!\!\!\!\!\!\! 
+ 2\left[ g \, a \, Q \left( \tau_2 \right) - p_2    +   \left(p_2-p_1 \right) \frac{g^2 a^2 Q^2(\tau_2)}{\vert {\vec p_1} + {\vec p_2} \vert^2 + 2 g^2 a^2 Q^2 \left( \tau_2 \right)}  \right] \, {\cal C}_+ \left( \tau_1 ,\, \tau_2 ;\, p_2 \right) \nonumber\\
&&  \!\!\!\!\!\!\!\!\!\!\!\!\!\!\!\!\!\!\!\!\!\!\!\!\!\!\!\!\!\!\!\!\!\!  \!\!\!\!\!\!\!\!  \times  \left\{ \left[ g \, a \, Q \left( \tau_1 \right) - p_2  +   \left(p_2-p_1 \right) \frac{g^2 a^2 Q^2(\tau_1)}{\vert {\vec p_1} + {\vec p_2} \vert^2 + 2 g^2 a^2 Q^2 \left( \tau_1 \right)}  \right] \,  \partial_{\tau_1} \, \partial_{\tau_2} {\cal C}_+ \left( \tau_1 ,\, \tau_2 ;\, p_1 \right) +  \partial_{\tau_1} \left[ g \, a \, Q \left( \tau_1 \right)  \right] \,   \, \partial_{\tau_2} {\cal C}_+ \left( \tau_1 ,\, \tau_2 ;\, p_1 \right) \right\} 
 \Bigg\} \Bigg]  \;,  \nonumber\\ 
\end{eqnarray} 
and where the symmetry $\vec{p}_1 \leftrightarrow \vec{p}_2$ has been exploited to write the last four lines in a more compact way. In terms of the rescaled variables defined in eq. (\ref{code-var}), recalling that $g a Q \left( \tau \right) = - \frac{m_Q}{\tau}$, and disregarding the slow roll variation of $m_Q$, ${\cal T}$ acquires the form (\ref{calT}) written in the main text. 

We then use the identity $\vert \vec{\epsilon}_L \left( {\hat p}_1 \right) \cdot  \vec{\epsilon}_L \left( {\hat p}_2 \right) \vert^2 = \frac{\left( {\hat p}_1 \cdot {\hat p}_2 - 1 \right)^2}{4} $ and note that the two $\delta-$functions in the second line of (\ref{deltaP-par2}) produce the same result. This gives 
\begin{eqnarray} 
& & \!\!\!\!\!\!\!\!  \delta \left\langle \delta \chi \left( \tau ,\, \vec{k}_1 \right) \,  \delta \chi \left( \tau ,\, \vec{k}_2 \right) \right\rangle' 
=  \frac{\lambda^2}{f^2}  \int^\tau d \tau_1 \, \int^{\tau_1} d \tau_2 
\int \frac{d^3 p_1 d^3 p_2}{\left( 2 \pi \right)^3} \, 
\frac{\left( {\hat p}_1 \cdot {\hat p}_2 - 1 \right)^4}{16}  \, \delta^{(3)}\left( \vec{k}_1 - \vec{p}_1 - \vec{p}_2  \right)  \, {\cal T} \nonumber\\  
& & \quad\quad \quad = 
\frac{\lambda^2}{f^2}  \int^\tau d \tau_1 \, \int^{\tau_1} d \tau_2 \, \frac{1}{64 \pi^2 k_1} \int_0^\infty d p_1 \, p_1 \, \int_{\vert k_1 - p_1 \vert}^{k_1 + p_1} d p_2 \, p_2 \, \left( \frac{k_1^2 - \left( p_1 + p_2 \right)^2}{2 p_1 p_2} \right)^4 \, {\cal T} \;. 
\end{eqnarray} 
To write the second line of this expression, we exploited the $\delta-$function, to have a three-dimensional integral over $d^3 p_1$. We then chose polar coordinates, with $\vec{k}_1$ oriented along the $z-$axis. The $\int d \phi$ integration is then trivial, and the $\int d \theta$ interaction can be traded for an integration over $p_2$.

Dividing by the linear power spectrum we then obtain the ratio reported in eq. 
(\ref{R-chi}) of the main text.

%%%%%%%%%%%%%%%%%%%%
\section{Semi-Analytic Approximation to Numerical Result}
\label{sec:appC}
%%%%%%%%%%%%%%%%%%%%

We devote this appendix to understanding the numerical results shown in section \ref{sec:nonlinear} via semi-analytical methods.  Schematically, the integral (\ref{R-chi}) that we want to evaluate has the form
\begin{eqnarray}
{\cal R}_{\delta \chi}  &=& {\hat N} \, \int dq_1 \, dq_2 \, dx_1  \, dx_2  \,\, {\cal P}(q_1,q_2) \cdot  \nonumber\\
&&Re\left\{ \frac{\left( X_c(x) X_c^*(x_1) - X_c^*(x) X_c(x_1) \right) \cdot \left( X_c(x) X_c^*(x_2) \right)}{\vert X(x) \vert^2}  \times \,\, \left( {\cal W}_{\rm sym}(x_1,x_2,q_1,q_2)\right) \right \} \;, \nonumber\\
\end{eqnarray}
where $q_2 \equiv \vert \hat{k}-\vec{q}_1 \vert$ , ${\cal W}_{\rm sym} \equiv \frac{{\cal W}(x_1,x_2,q_1,q_2) + {\cal W}(x_1,x_2,q_1,q_2)}{2}$, where 
\begin{equation}
{\cal P} \left( q_1,\,q_2 \right) \equiv \left(\frac{q_1^2+q_2^2+2 q_1 q_2 -1}{2 q_1 q_2}\right)^4 \;, 
\end{equation} 
is the prefactor resulting from polarization vectors, and 
\begin{equation}
{\hat N} \equiv \frac{\lambda^2 \, H^2 \, \left(1+m_Q^2 \right)}{256\pi^2  f^2} \;, 
\end{equation} 
is the constant factor in front of the integral. 

At first, we observe that since the first factor inside the curly bracket is purely imaginary and the denominator is purely real, the final contribution to the integral will come from two options: imaginary part of the axion mode function product and real part of ${\cal W}_{\rm sym}$ or real part of the axion mode function product and the imaginary part of ${\cal W}_{\rm sym}$. Through a numerical evaluation, we verified that the first option dominates over second one. Then the final result acquires the form 
\begin{eqnarray}
{\cal R}_{\delta \chi}  &=& {\hat N} \, \int dq_1 \, dq_2 \, dx_1  \, dx_2  \,\, {\cal P}(q_1,q_2) \cdot  \nonumber\\
&&\!\!\!\!\!\!\!\!\!\!\!\!\!\!\!\!
\left\{ \frac{\left( X_c(x) X_c^*(x_1) - X_c^*(x) X_c(x_1) \right) \cdot \left( X_c(x) X_c^*(x_2) - X_c^*(x) X_c(x_2) \right) /2 }{\vert X(x) \vert^2}  \times \,\,Re \left[{\cal W}_{\rm sym}(x_1,x_2,q_1,q_2)\right] \right\} \;. \nonumber\\
\end{eqnarray}

We further observe that ${\cal W}_{\rm sym}$ leads to a  symmetric and  separable structure between $x_1$ and $x_2$; therefore the domain of the $\int d x_1 d x_2$ integration can be made rectangular  \footnote{Specifically, we can write $\int_a^b d x_1 \int_{x_1}^b d x_2 f(x_1,x_2) = \frac{1}{2} \int_a^b d x_1 \int_a^bd x_2 f(x_1,x_2) $ since the integrand $f$ is symmetric.}. Keeping this into account, we see that the integral can be recast in the form 
\begin{equation}
{\cal R}_{\delta \chi}  = {\hat N}  \int dq_1 \, dq_2 {\cal P}(q_1,q_2) \,\, \bigg\vert  \int   d x_1 \, G(x,x_1) \, S(q_1,q_2, x_1)  \bigg\vert^2  \;, 
\label{ininredgreen}
\end{equation}
where
\begin{eqnarray}
G(x,x_1) &=&  \frac{Im\left[ X(x)X^*(x_1) \right]}{\vert X(x) \vert}  \;, 
\end{eqnarray}
and 
\begin{eqnarray} 
S(q_1,q_2, x_1)  &\equiv& S_1(q_1,q_2, x_1)+S_2(q_1,q_2, x_1) 
\end{eqnarray}

With $S_1$ and $S_2$ defined as

\begin{eqnarray} 
S_1(q_1,q_2, x_1)  &\equiv& \frac{1}{x_1} \Bigg[ q_1 x_1 (-m_Q + q_2 x_1) \, t_c(q_2 x_1) \, t_c'( q_1 x_1) \nonumber\\ 
&& \quad\quad 
+   t_c(q_1 x_1) \left(m_Q  \, t_c(q_2 x_1) + q_2 x_1 (-m_Q + q_1 x_1) \,  t_c'(q_2 x_1)\right)  \Bigg] \;. \nonumber\\
S_2(q_1,q_2, x_1)  &\equiv & \frac{m_Q^2 (q_1-q_2)\,x_1\,\left[q_1\, t_c(q_2 x_1) t_c'(q_1 x_1)-q_2\, t_c(q_1 x_1)t_c'(q_2 x_1)\right]}{2 m_Q^2+x_1^2}
\end{eqnarray}
The source has been separated into two parts labeled as $S_1$ and $S_2$ because the time integral in (\ref{ininredgreen}) can be done analytically if we disregard the contribution of $S_2$. We have confirmed with the numerical results presented in the main text that the $S_2$ term can give corrections up to ${\cal O}(10\%-40\%)$ 
in the range of $m_Q$ that we have studied, with an increasing relative contribution at the largest values of  $m_Q$, for which the ratio ${\cal R}_{\delta \chi} $ is greater than one. This observation is supported by figure (\ref{fig:mQ-R}) where the dashed line was calculated using only $S_1$ whereas the data points were produced numerical using the full expression. For this reason we include only $S_1$ in the present semi-analytical estimate.  

As suggested by the notation, the formal result (\ref{ininredgreen}) is the one that is obtained solving the equation of motion for $\delta \chi$  in the presence of the source $\propto t$ via the Green function method. 

\begin{figure}[tbp]
\centering 
\includegraphics[width=0.5\textwidth,angle=0]{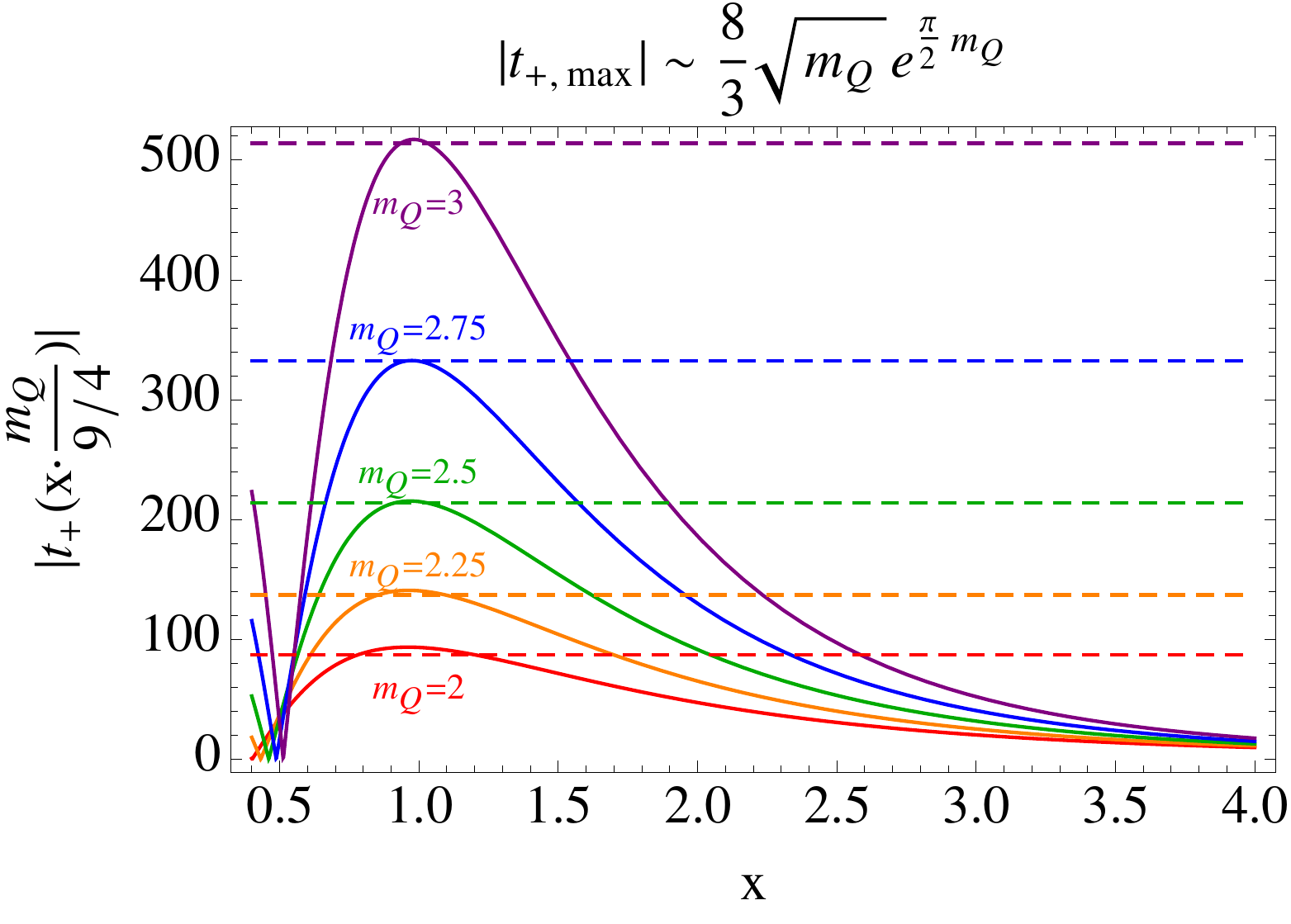}
\caption{The solid lines show the  exact evolution of the mode functions $t_+$ (eq. (\ref{exacttmode})) for different values of $m_Q$ (we note that time flows from right to left in the figure, since $x$ is proportional to $- \tau$). The dashed lines show the approximate relation  (\ref{tmaxapprox}) for the maximum amplitude. The argument of the mode function has been rescaled in a way proportional to $m_Q$,  so that the peak appears at the same horizontal position for all the cases shown in the figure.}
\label{fig:tmax}
\end{figure}

We verified that the maximum contribution to the integral above is coming from the bump of the tensor mode function $t_c$ that takes place around horizon crossing (more specifically, when the quantity in parenthesis in eq. (\ref{eq-t}) is negative). The behavior of $t_c$ is shown in the solid curves of Figure \ref{fig:tmax}. For the purpose of our computation, we can approximate the bump as 
\begin{equation}
t_c\left(x\right)=T_{max} \cdot \exp \left[- \frac{\log^2 \left(\frac{ x }{ \mu} \right)}{2\sigma^2}  \right] \,\, , \qquad  \mu \equiv \frac{4}{9} m_Q  \,\,  , \qquad \frac{1}{2 \sigma^2}\equiv m_Q \,\, ,
\label{tfitting}
\end{equation}

The parameter $\mu$ was chosen so that the approximated functions peak for $x= \frac{m_Q}{2.25}$. This is demonstrated in figure \ref{fig:tmax}. The value of $\sigma^2$ was chosen so to match the width obtained from the numerical evolution.  Comparing with the numerical solutions shown in Figure \ref{fig:tmax}, we see that  the maximum value of the mode function can be well approximated by 
\begin{equation}
\vert \, T_{max} \, \vert \simeq \frac{8}{3} \sqrt{m_Q}\exp\left(\frac{\pi}{2}m_Q\right) \;. 
\label{tmaxapprox}
\end{equation}
We checked the accuracy of (\ref{tfitting}) numerically; namely, we verified that the integral that produces ${\cal R}_{\delta \chi}$ with the approximated form (\ref{tfitting}) for the mode functions accurately reproduces the integral obtained with the exact $t_c$. Using (\ref{tfitting}), the function $S_1(q_1,q_2, x_1)$  becomes
\begin{eqnarray}
S_1(q_1,q_2, x_1)  &=& \frac{ T^{2}_{max} \, m_Q } { x_1 } \, e^{-m_Q\left(\log\left(\frac{9 q_1 x_1}{4 mQ}\right)^2+\log\left(\frac{9 q_2 x_1}{4 mQ}\right)^2\right)} \nonumber\\
& \times & \left(1+2\left(m_Q-q_2 x_1\right)\log\left(\frac{9 q_1 x_1}{4 m_Q}\right)+2\left(m_Q-q_1 x_1\right)\log\left(\frac{9 q_2 x_1}{4 m_Q}\right)\right) \;. 
\end{eqnarray}
Studying the numerical solutions for the scalar mode function $X$, we then find that the combination  $\frac{Im\left(X(x)X^*(x_1)\right)}{\sqrt{\vert X(x) \vert^2}}$ is well approximated by 
\begin{equation}
\frac{Im\left(X(x)X^*(x_1)\right)}{\sqrt{\vert X(x) \vert^2}} \bigg\vert_{x \to 0}\simeq \frac{1}{\bf \Lambda}\left(0.22-\frac{0.34}{m_Q}\right) \, x_1^2 \;. 
\label{chifitting}
\end{equation}

Using the approximations (\ref{chifitting}) and (\ref{tfitting}) one can calculate the integral over $x_1$ in the expression (\ref{ininredgreen}) analytically:  
\begin{eqnarray}
{\cal R}_{\delta \chi}  &\simeq & {\hat N} \frac{\vert T_{max}\vert^4}{{\bf \Lambda}^2 } \left(0.22-\frac{0.34}{m_Q}\right)^2 \, \int dq_1 \, dq_2 \,{\cal P}(q_1,q_2) \, \bigg[  \frac{16}{729\,q_1^3} \, e^{-m_Q\log\left(\frac{q_1}{q_2}\right)^2}m_Q^{5/2} \sqrt{2\pi}  \hfil \nonumber\\
 &&  \!\!\!\!\!\!\!\!\!\!\!\!\!\!\!  \times \bigg\{  \frac{27}{2}\,q_1\,e^{\frac{\left(1+m_Q\log\left(\frac{q_1}{q_2}\right)\right)^2}{2 m_Q}}+ e^{\frac{\left(3+2m_Q\log\left(\frac{q_1}{q_2}\right)\right)^2}{8 m_Q}}\left(-3\left(q_1+q_2\right)+2 m_Q\left(q_1-q_2\right)\log\left(\frac{q_1}{q_2}\right)\right)  \bigg\}  \bigg]^2 \;.  \qquad  \qquad  
\label{integrandqq}
\end{eqnarray}

\begin{figure}[tbp]
\centering 
\includegraphics[width=0.6\textwidth,angle=0]{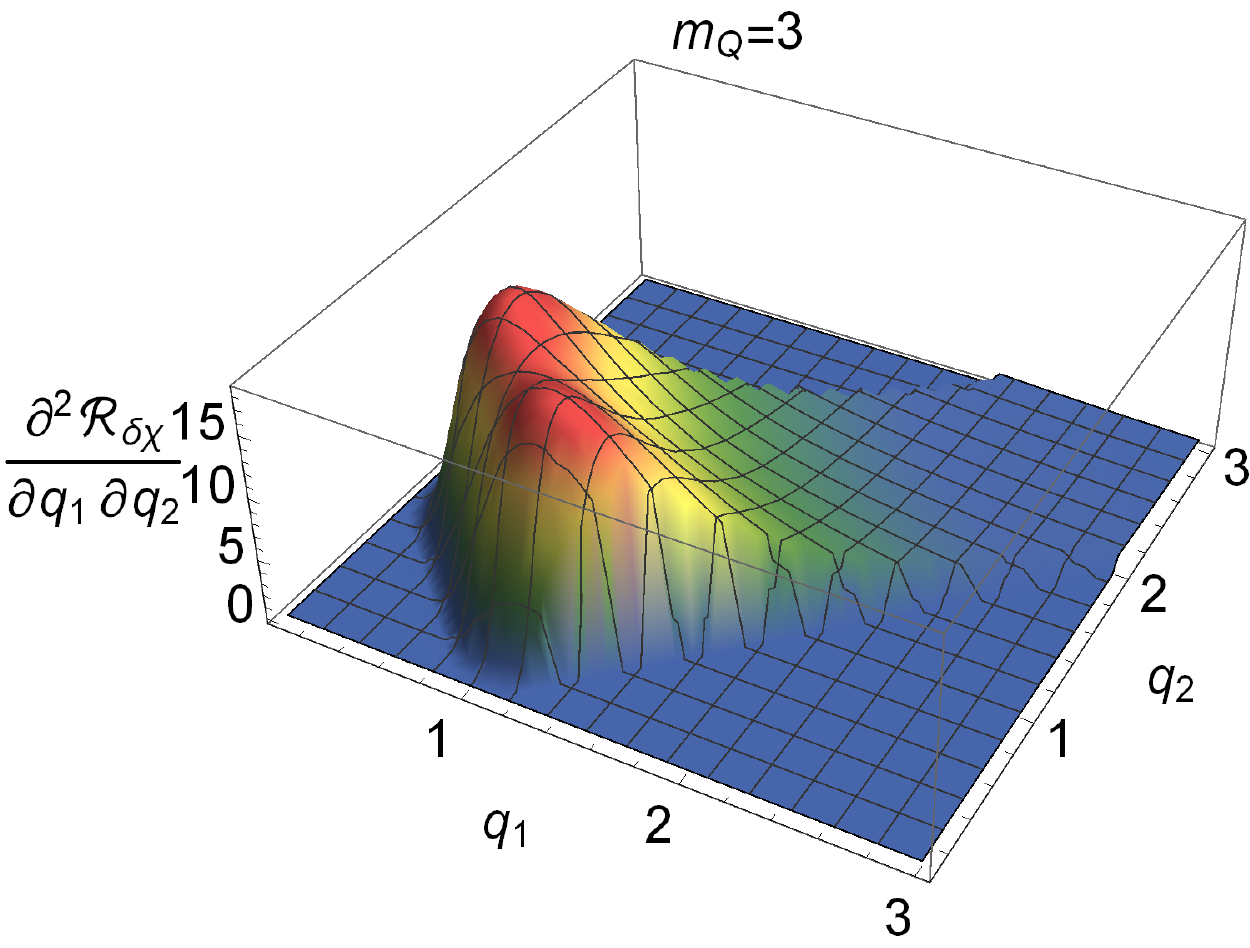}
\caption{$3$ dimensional plot of the integrand of (\ref{integrandqq}), for a specific choice of parameters leading to $m_Q = 3$. 
} 
\label{fig:integrantq1q2}
\end{figure}

The integrand entering in this expression is shown in Figure \ref{fig:integrantq1q2}. We see that this expression is peaked at $q_i = {\rm O } \left( 1 \right)$, namely when the momenta $p_i$ of the sourcing modes $t$ are comparable to the momentum of the mode function $\delta \chi$ (recall that $q_i \equiv \frac{p_i}{k}$). This is analogous to what happens in the Abelian case where the inflaton perturbations are sourced by the vector modes amplified by a $\chi F {\tilde F}$ coupling \cite{Barnaby:2010vf}. 

The results of the double integral above are calculated for varying $m_Q$ ranging from $1.8$ to $3$ and the result is fitted as
\begin{equation}
{\cal R}_{\delta \chi}  \simeq {\hat N} \frac{|T_{max}|^4}{{\bf \Lambda}^2} \left(0.22-\frac{0.34}{m_Q}\right)^2 \left( 11.3 + 1.48\cdot m_Q^{5} \right) \;. 
\end{equation}
Next, we use the fact that $m_Q=\frac{g Q}{H}$ and that ${\bf \Lambda} =\frac{\lambda Q}{f}$ to  rewrite ${\hat N}$ in terms of $m_Q$ and ${\bf \Lambda}$: 
\begin{equation}
{\hat N} = \frac{\lambda^2 \, H^2 \, \left(1+m_Q^2 \right)}{256\pi^2  f^2}= \frac{g^2 {\bf \Lambda}^2 \left(1+m_Q^2\right)}{256 \pi^2 m_Q^2} \;, 
\end{equation}
so that the explicit dependence of ${\cal R}_{\delta \chi}$ on ${\bf \Lambda}$ drops out. We then insert the explicit expression (\ref{tmaxapprox}) for $\vert T_{\rm max} \vert$, so to obtain   
\begin{eqnarray}
{\cal R}_{\delta \chi} &\simeq& \frac{1}{256 \pi^2} \frac{g^2(1+m_Q^2)}{m_Q^2} \left(0.22-\frac{0.34}{m_Q}\right)^2  \left( \frac{8}{3} \,\,  \sqrt{m_Q} \,\,  e^{\frac{\pi}{2} m_Q}  \right)^4 \left( 11.3 + 1.48\cdot m_Q^{5} \right) \;. \nonumber\\
\label{Rincomplete}
\end{eqnarray}

We now show that the value of $g$ can be related to $m_Q$ along the inflationary trajectory. We verified that the super-horizon evolution of the mode function of the inflaton perturbation is well fitted by 
\begin{equation}
\vert x\sqrt{2k} \hat{X} \vert\simeq \frac{1+m_Q^2}{\bf \Lambda}\left(\frac{-12.5}{m_Q} + \frac{255}{m_Q^2}  - \frac{897}{m_Q^3} +  \frac{1050}{m_Q^4}\right) \;. 
\label{superhorizon}
\end{equation} 
Using the slow roll relations (\ref{slowroll}) and $H \simeq \frac{M_p}{\sqrt{3}} {\tilde \mu^2} \sqrt{1+\cos {\tilde x}}$, 
we can write 
\begin{equation}
P_\zeta^{(0)} \equiv \frac{k^3}{2 \pi^2} \vert \zeta \vert^2 = 
\frac{1+\cos {\tilde x}_N}{48 \pi^2} \, \frac{\lambda^2 \, {\tilde \mu}^4}{{\tilde f}^2} \, \frac{m_Q^2}{\left( 1 + m_Q^2 \right)^2} \, 
\left\vert x \; \sqrt{2 k} {\hat X} \right\vert^2  \;. 
\end{equation} 
We insert the expression (\ref{superhorizon}) for the last factor, with ${\tilde x}$ given by eq. (\ref{xN}). We then use the first approximate equality in (\ref{Lambda-scaling}) for ${\bf \Lambda}$. Finally we set $P_\zeta^{(0)}=2.2\cdot 10^{-9}$ according to the observed value \cite{Ade:2015lrj}. This gives 
\begin{equation}
\tilde{\mu}\simeq 7.42\cdot 10^{-4} \sqrt{\lambda} \frac{\left(1+m_Q^2\right)^{1/4}}{m_Q^{3/2}}\left(\frac{-12.5}{m_Q} + \frac{255}{m_Q^2}  - \frac{897}{m_Q^3} +  \frac{1050}{m_Q^4}\right)^{-1/2} \;. 
\end{equation}
After substituting this relation in (\ref{mQ-N}), and setting $N=60$, we obtain 
\begin{equation}
g(m_Q) \simeq \frac{5.8 \cdot 10^{-4} \sqrt{1+m_Q^2}}{ \frac{-12.5}{m_Q} + \frac{255}{m_Q^2}  - \frac{897}{m_Q^3} +  \frac{1050}{m_Q^4} } \;. 
\label{eq:gmq}
\end{equation} 
We insert the relation (\ref{eq:gmq}) in eq. (\ref{Rincomplete}), and we fit the expression ${\cal R}_{\delta \chi} / {\rm e}^{2 \pi m_Q} $ with a monomial $D \, m_q^\gamma$ in the $1.8 \leq m_Q \leq 3$ range. We obtain 
\begin{eqnarray}
{\cal R}_{\delta \chi} & \simeq & 6.6 \cdot 10^{-16}\, m_Q^{18}  \, e^{2 \pi m_Q} \;. 
 \label{Rmqestfinal}
\end{eqnarray}

This expression agrees  well with the result of the numerical integration, as seen in Figure \ref{fig:mQ-R}.

\end{document}